\documentclass[aps,prl,twocolumn,longbibliography,superscriptaddress]{revtex4-2}
\usepackage{amsfonts}
\usepackage{amsmath}
\usepackage{graphicx}
\usepackage{dsfont}
\usepackage{diagbox}
\usepackage{array}
\usepackage{amssymb}
\usepackage{bbm}
\usepackage{float}
\usepackage{subfigure}
\usepackage{MnSymbol}
\usepackage{url}
\usepackage{times}
\usepackage{manfnt}
\usepackage{booktabs}
\usepackage[colorlinks=true,
            linkcolor=blue,
            anchorcolor=blue,
            urlcolor=blue,
            citecolor=blue
            ]{hyperref}
\usepackage{xcolor}
\pagecolor{white}
\usepackage{ulem}
\usepackage{physics}
\usepackage{enumitem}
\usepackage{wasysym}

\newcommand\redsout{\bgroup\markoverwith{\textcolor{red}{\rule[0.5ex]{2pt}{0.8pt}}}\ULon}

\begin{document}

\title{Topological Fracton Quantum Phase Transitions by Tuning Exact Tensor Network States}

\author{Guo-Yi Zhu}
\email{gzhu@uni-koeln.de}
\affiliation{Institute for Theoretical Physics, University of Cologne, Z\"ulpicher Straße 77, 50937 Cologne, Germany}

\author{Ji-Yao Chen}
\email{chenjiy3@mail.sysu.edu.cn}
\affiliation{Guangdong Provincial Key Laboratory of Magnetoelectric Physics and Devices, School of Physics, Sun Yat-sen University, Guangzhou, 510275, China}
\affiliation{Dahlem Center for Complex Quantum Systems, Freie Universit\"at Berlin, 14195 Berlin, Germany}

\author{Peng Ye}
\email{yepeng5@mail.sysu.edu.cn}
\affiliation{Guangdong Provincial Key Laboratory of Magnetoelectric Physics and Devices, School of Physics, Sun Yat-sen University, Guangzhou, 510275, China}
\affiliation{State Key Laboratory of Optoelectronic Materials and Technologies, Sun Yat-sen University, Guangzhou, 510275, China}

\author{Simon Trebst}
\email{trebst@thp.uni-koeln.de}
\affiliation{Institute for Theoretical Physics, University of Cologne, Z\"ulpicher Straße 77, 50937 Cologne, Germany}

\date{\today}

\begin{abstract}
Gapped fracton phases of matter generalize the concept of topological order and broaden our fundamental understanding of entanglement in quantum many-body systems. However, their analytical or numerical description beyond exactly solvable models remains a formidable challenge. Here we employ an exact 3D quantum tensor network approach that allows us to study a $\mathbb{Z}_N$ generalization of the prototypical X cube fracton model
and its quantum phase transitions between distinct topological states via fully tractable wave function deformations. We map the (deformed) quantum states exactly to a combination of a classical lattice gauge theory and a plaquette clock model, and employ numerical techniques to calculate various entanglement order parameters. For the $\mathbb{Z}_N$ model we find a family of (weakly) first-order fracton confinement transitions that in the limit of $N\to\infty$ converge to a continuous phase transition beyond the Landau-Ginzburg-Wilson paradigm. We also discover a line of 3D conformal quantum critical points (with critical magnetic flux loop fluctuations) which, in the $N\to\infty$ limit, appears to coexist with a gapless deconfined fracton state.

\end{abstract}

\maketitle


%
Quantum states with intrinsic topological order distinguish themselves through long-range entanglement \cite{Wen17zoo}, quasiparticle excitations with   
exotic statistics \cite{Nayak08rmp}, and their applicability as quantum memories \cite{Kitaev01toriccode}.
Such states have been widely studied in two spatial dimensions (2D), e.g.\ as ground states of the toric code (TC) \cite{Kitaev01toriccode}, which has also allowed for recent experimental realizations engineered in state-of-the-art quantum simulators \cite{Roushan21sycamoretoriccode, Lukin21toriccode}.
Exploring such states in three-dimensional (3D) settings has given rise to the family of {\sl fracton} topological orders \cite{Chamon05fracton, Haah11, Yoshida13, Fu16} with strictly immobile excitations, the eponymous {\sl fractons}, which have ignited interest not only in the fields of quantum information and quantum matter but also elasticity and gravity~\cite{Hermele19reviewfracton,You20fractonreview}.
The simplest exactly solvable fracton model is the X cube (XC)~\cite{Fu16} where the mobility constraint is deeply rooted in the absence
of string operators \cite{Haah11}. 
The ground states of the XC span a degenerate manifold 
which is insensitive to local perturbations and 
whose peculiar subextensivity can be traced back to an intimate connection to 2D topological order 
via a coupled-layer-construction~\cite{Hermele17, Vijay17coupledLayersXcube, Shirley2018, Chen20coupledCS, Aasen20tqftnetwork, Slagle21foliatedfield, LiYe20fracton, LiYe21fracton}.
Like the TC which is equivalent to a $\mathbb{Z}_2$ lattice gauge theory~\cite{Kitaev01toriccode}, the XC can be viewed as a generalized lattice gauge theory coupled to  $\mathbb{Z}_2$ matter  with certain subsystem symmetries~\cite{Fu16}; in the long wave-length limit, it is equivalent to an off-diagonal $U(1)$ tensor gauge theory that is turned massive via a Higgs mechanism~\cite{XuWu08RVP, Pretko17tensorgauge, Chen18tensorhiggs, Barkeshli18tensorhiggs, Seiberg21ZNfracton} 
and where the matter charge has conserved higher moments  \cite{Pretko18fractongaugeprinciple, Gromov19multipoleAlgebra, Seiberg19vectorSym, YuanYe20fractonSF, ChenYe21fractonSF, LiYe21fractonRG, YuanYe22fracton, Nandkishore21sym, Ye23fractonSF}.

Despite this impressive understanding of fracton physics, there are still a number of unresolved questions. One is the principal nature of quantum phase transitions (QPTs) involving fracton topological phases, which due to their nonlocal structure have to go beyond the traditional Landau-Ginzburg-Wilson paradigm. 
Although there have been analytical attempts 
based on Hamiltonian duality, series expansions or phenomenological field theories~\cite{Vijay17coupledLayersXcube, Kim17, Sondhi18fractondiagnostics, Schmidt20fractonpcut, Liu21fracton, Hermele21fracton, Williamson21SubdimensionalCriticality, Walther22Xcube}, a direct microscopic investigation, e.g., by considering deformations of exactly solvable Hamiltonians as for their 2D counterparts, has remained largely out of scope of current approaches. 

Here we follow a different route and study {\sl wave function deformations} that allow us to move from the exactly known ground states of certain fracton models through QPTs to topologically trivial states devoid of any fractons. 
Such paths differ from the conventional {\sl Hamiltonian deformations} by spacetime anisotropy and could yield space-conformal quantum critical points \cite{Fradkin04RK}. 
In doing so, we employ tensor network (TN) wave functions that allow us to {\sl exactly} tune these quantum states along a chosen path -- an approach previously employed in the context of 2D topological order \cite{Cirac21rmp, Fradkin04RK, Troyer10topocrit, Verstraete13toriccode, Verstraete15shadow, Schuch17anyoncondensates, Chen18z2rvb, Zhu19ToricCode, gmz20fibonacci, gmz20doublesemion}. 
For the 3D fracton order of interest here, we identify QPTs along the path by numerical TN calculations that are based on an 
analytical quantum-classical mapping 
and  allow us to calculate various 
entanglement order parameters~\cite{Schuch21entanglementorder, Verstraete13toriccode, Verstraete15shadow, Poiblanc17criticalpeps, Corboz18, Lauchli18, Zhu19ToricCode, Chen20su3peps} as diagnostics. 
For the $\mathbb{Z}_N$ generalized XC fracton model  our main findings can be concentrated around
a line of \textit{3D space-conformal deconfined quantum critical points}, where the deconfined fractons are connected by critical fluctuating strings. 
In one direction, the string can either condense resulting in a gapped $\mathbb{Z}_N$ fracton phase, or confine a pair of fractons into a fracton dipole thereby falling into the stacked 2D TCs phase; 
in the other direction, monopole proliferation leads to a complete fracton confinement in a weak first-order transition that turns {\sl continuous} in the unHiggsed $N\to\infty$ limit.

\begin{figure}
    \centering
    \includegraphics[width=\columnwidth]{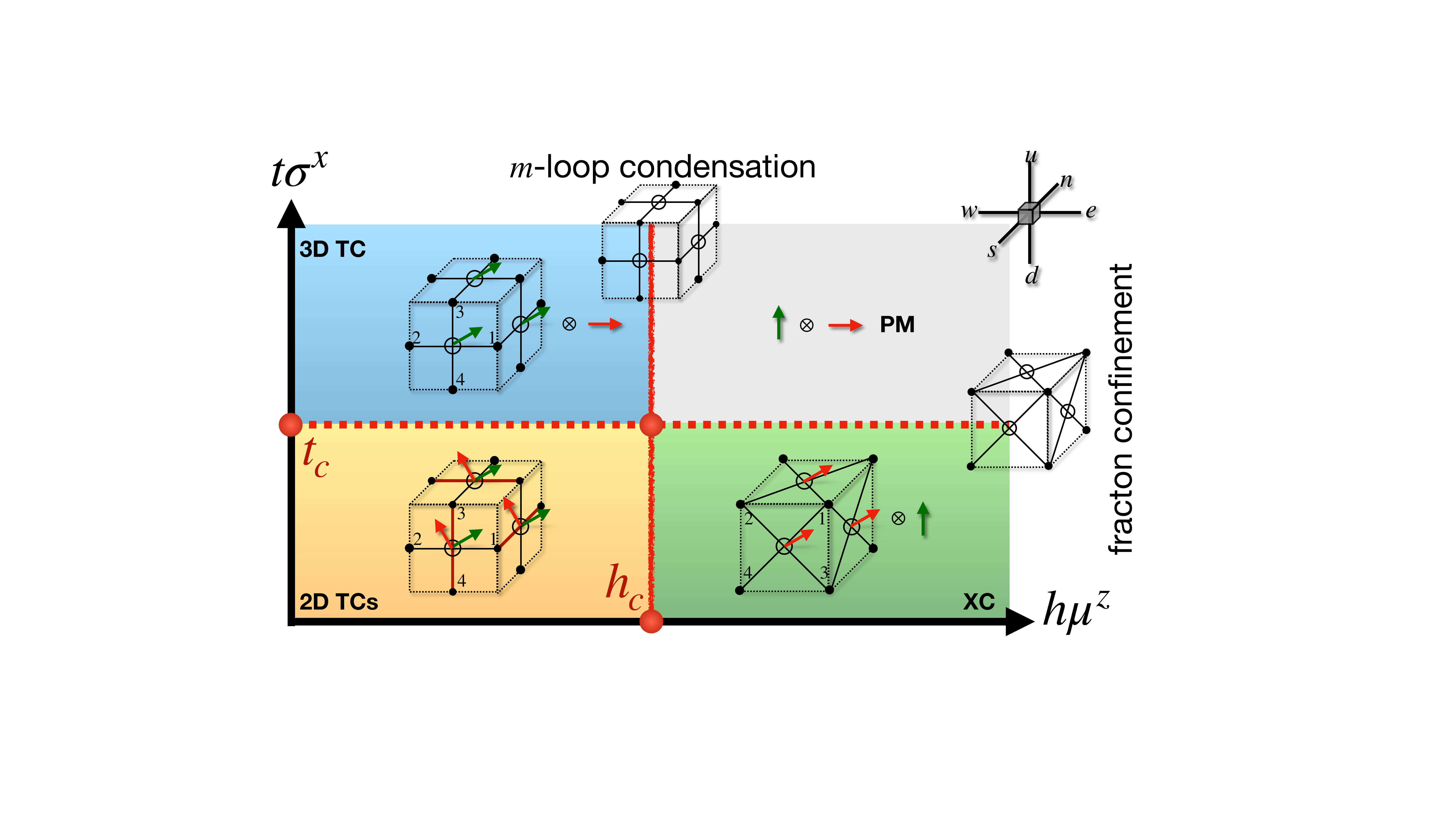}
    \caption{{\bf Phase diagram of tuning the exact quantum TN state} 
    between the stacked 2D toric codes, 3D toric code, X cube, and the trivial paramagnet (PM) for $\mathbb{Z}_N$ gauge group. 
    The TN states are illustrated in dual cubes, where the black dots denote the virtual variables and the red (green) arrow denotes physical spin $\sigma^z(\mu^z)$, satisfying $\mu^z=\omega^{n_1-n_2-n_3+n_4},\ \sigma^z=\omega^{n_4-n_3}$.
    The classical TNs lying at the QPTs show the wave function norm. 
 }
    \label{fig:phasediagram}
\end{figure}

\textit{Parent wave function.--}
Our starting point is the observation that stacked intersecting 2D TCs can act as a parent
for realizing both the XC and 3D TC models 
through condensing its elementary excitations -- magnetic flux loops or
monopoles~\cite{Hermele17, Vijay17coupledLayersXcube, Kim17, Vijay21hybridfracton}.
In terms of wave functions, this motivates us to adopt the ground state of the stacked 2D TCs as a parent wave function and study its deformations that will pass through QPTs to either the XC or 3D TC ground states, as illustrated in Fig.~\ref{fig:phasediagram}. At the fixed point the wave function is a stack of dual 2D Ising quantum paramagnets \cite{Kitaev01toriccode}
\begin{equation}
    |\psi_0\rangle
    = \sum_{\{s\}} |Z_{p\cap q}=s_p s_q\rangle \,,
    \label{eq:psi0}
\end{equation}
where the Pauli $z$ matrix $Z$ for the physical spin on a link $l=p\cap q$ can be recast as the domain wall between classical Ising spins $s=\pm 1$ on the two adjacent plaquettes $p,q$ in the same plane. On every link this results in two spins from two intersecting planes.
Now let us rotate the local basis into
$
    \mu_l^z\equiv Z_{l_1}Z_{l_2},\ \mu_{l}^x\equiv X_{l_2},\ 
    \sigma_{l}^x\equiv X_{l_1}X_{l_2},\ \sigma_{l}^z\equiv Z_{l_1},
$
where $l_1,l_2$ lie in $xz(yx)(zy)$ and $xy(yz)(zx)$ planes, respectively (see Supplemental Material (SM)~\cite{supplemental}). 
The ground state fulfills
\begin{equation}
    \begin{split}
        \prod_{l\in \mbox{\varhexstar}}\mu_{l}^z=1,\quad
        \prod_{l\in\ \mbox{\mancube}}\sigma_{l}^x=1 \,
        \label{eq:Gausslaw}
    \end{split}
\end{equation}
for every vertex $\varhexstar$ and cube $\mbox{\mancube}$, respectively. 
These are the Gauss laws for the 3D vector and tensor gauge theories, respectively, 
which makes it natural to interpret the subsystem $\{\mu^z\}$ as the 3D TC~\cite{Wen053dtoriccode}, while $\{\sigma^x\}$ as the XC~\cite{XuWu08RVP, Fu16, Pretko17tensorgauge}.
The parent state $|\psi_0\rangle$ is free of charge, while any violations of Eq.~\eqref{eq:Gausslaw} 
define the boson ($e$) and fracton ($f$) charge excitations. 
The magnetic flux loop ($m$-loop) is composed of a loop of $m$-particles that penetrate plaquettes with $\prod_{l\in\square} \mu_l^x=-1$~\cite{Wen053dtoriccode}, while the magnetic monopole is defined on the site satisfying $\prod_{l\in +}\sigma_l^z = -1$ for inplane vertices $+$~\cite{Fu16}. 
Notice that the monopoles in the XC subsystem are entangled with the proliferating electric string turning points of the TC subsystem. Thus if either one subsystem is traced out, one would get a mixed state with $m$-loops or monopole excitations~\cite{supplemental}. 

To explore the QPTs induced by condensing these elementary excitations, we add fluctuations of the $m$-loop and monopoles by a local nonunitary deformation
\footnote{
Note that we can construct an exact Rokhsar-Kivelson type, frustration free Hamiltonian~\cite{Henley04RK, Fradkin04RK} that stabilizes 
\mbox{$|\psi(t,h)\rangle$} as its ground state
 $   H_{\rm RK} = H_0 + \sum_{\square}\hat{V}_{\square} + \sum_+\hat{V}'_{+}$,
    %
with $H_0$ being the Hamiltonian of stacked uncoupled 2D TCs, while $\hat{V}_{\square} = e^{-h\sum_{l\in\square}\mu_l^z }$ and $\hat{V}'_{+} = e^{-t\sum_{l\in +} \sigma_l^x }$.
A detailed proof is given in the SM. 
}
\begin{equation}
|\psi(t,h)\rangle = \exp\left({\frac{1}{2} \sum_{l} h \mu_{l}^z + t\sigma_{l}^x}\right) |\psi_0\rangle\,.
    \label{eq:psith}
\end{equation}
Here $h\mu^z$ fluctuates $m$-loops and adds electric string tension to confine the boson charge, distilling the XC state from $|\psi_0\rangle$, which in the $h\to\infty, t=0$ limit turns into the exactly solvable XC TN state as a cuboid condensate~\cite{supplemental, Bernevig18xcube, Schuch21stability}. 
$t\sigma^x$ fluctuates the monopole (lineon) and turns on electric membrane tension to confine the fracton, distilling the 3D TC out of $|\psi_0\rangle$ as a loop condensate.
Such deformations can be captured by an exact, frustration-free  Rokhsar-Kivelson Hamiltonian, which in the perturbative limit is equivalent to turning on magnetic fields along $\mu^z$ and $\sigma^x$~\cite{supplemental}. 
In our numerical analysis, we express the state \eqref{eq:psith} as a 3D projected entangled paired state (PEPS) with finite bond dimension (see Fig.~\ref{fig:phasediagram}).


\textit{Quantum classical correspondence.--}
The TN wave functions \eqref{eq:psith} can be mapped onto effectively classical models by defining $\langle \psi |\psi\rangle$ as a partition function~\cite{Henley04RK, Fradkin04RK, Isakov2011, Zhu19ToricCode} over the ensemble of the virtual TN  variables. We find analytically that this partition function {\sl factorizes} into
\begin{equation}
	\label{eq:factorization}
	\langle \psi | \psi \rangle = \sum_{\{s\}} e^{-\epsilon_g} \times \sum_{\{\tau\}} e^{-\epsilon_p} \,,
\end{equation}
where the two terms
\begin{equation}
\includegraphics[width=.6\columnwidth]{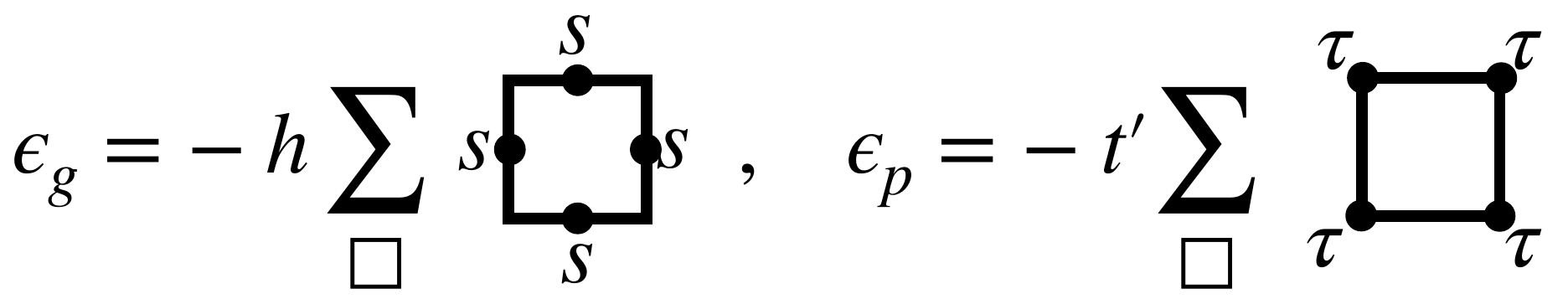}
\label{eq:psithClassicModel}
\end{equation}
are precisely the  cubic lattice variants of the classical $\mathbb{Z}_2$ gauge~\cite{Kogut79rmp} and plaquette~\cite{Wegner94PIM, Wegner98PIM, MooreXu04PIM} models, describing the (fluctuating) loop gas and fracton matter, respectively. 
Here $s$ creates the $m$-particle~\cite{Zhu19ToricCode}, while $\tau$ measures the fractons in the dual lattice~\cite{supplemental}. 
For the latter, the parameter $t'\equiv \frac{1}{2}\ln\coth t$ is obtained via a Kramers-Wannier relation \cite{Johnston14PIM}.
An immediate consequence of this factorization is that 
the phase diagram in Fig.~\ref{fig:phasediagram} is controlled by two independent QPTs, tuned by $h$ and $t$ respectively. 
The physical origin is that the $m$-loops and the monopoles have trivial mutual statistics and can thus condense simultaneously. 
The classical TNs for $\epsilon_g$ and $\epsilon_p$ can both be equivalently generated, through cube-vertex duality, by a rank-6 tensor (Fig.~\ref{fig:phasediagram}, top right) where the tensor bond takes the plaquette variable $W_\square=\prod_{l\in\square} s_l$ or $W_\square'=\prod_{j\in\square} \tau_j$. 
To contract the TNs $\langle \psi | \psi \rangle$, 
we  employ the variational infinite projected-entangled-paired-state(iPEPS) method to get the dominant boundary fixed point~\cite{Verstraete16gradientmethod, Corboz16ipeps, Verstraete18classic3D, Xiang19ADtensor, Corboz22ipeps3d, Wright06optimBook} (see SM). 
As in general PEPS, the virtual variables serve as the entanglement degrees of freedom that are responsible for stretching out the long range entanglement~\cite{Cirac21rmp,Verstraete13toriccode, Verstraete15shadow, Zhu19ToricCode, Schuch21entanglementorder}.
This allows us to set up a dictionary between the quantum correlation and the classical correlations as in Table~\ref{tab:quantumclassicalcorrespondence}.
We elucidate the nature of the two QPTs in the following.  

\begin{table}[t]
\centering
\caption{{\bf Quantum classical correspondence} between wave function and statistical model.
}
\label{tab:quantumclassicalcorrespondence}
\begin{tabular}[t]{c  | c  | c}
\toprule
Quantum toric code& 
Tensor network  &
Classical gauge  \\
\midrule
$\mu^z$ &
$Z_\square$ &
$W_\square$\\
$\langle e|e\rangle$ & 
$X_\square$ &
't Hooft string \\
$\langle \psi | \prod_{p\in \partial M} m_p \rangle $ & 
$\prod_{\square\in M} Z_\square$&
Wilson loop \\
\bottomrule
\toprule
Quantum fracton& 
Tensor network &
Classical plaquette \\
\midrule
Quadrupole & 
$Z_\square$ &
$W_\square'$\\
$\langle \psi | \text{monopole} \rangle$ & 
$X_\square$ &
Twist defect \\
$\norm{\ket{\prod_{j\in\partial\partial M} f_j}}^2$& 
$\prod_{\square\in M} Z_\square$&
$\prod_{j\in\partial\partial M} \tau_j$ \\
$-\ln\langle \psi | \psi \rangle$ &
$-\ln \text{tTr}\prod_j \hat{T}(j)$ &
Free energy\\
\bottomrule
\end{tabular}
\end{table}

\begin{figure}[t]
    \centering
    \includegraphics[width=\columnwidth]{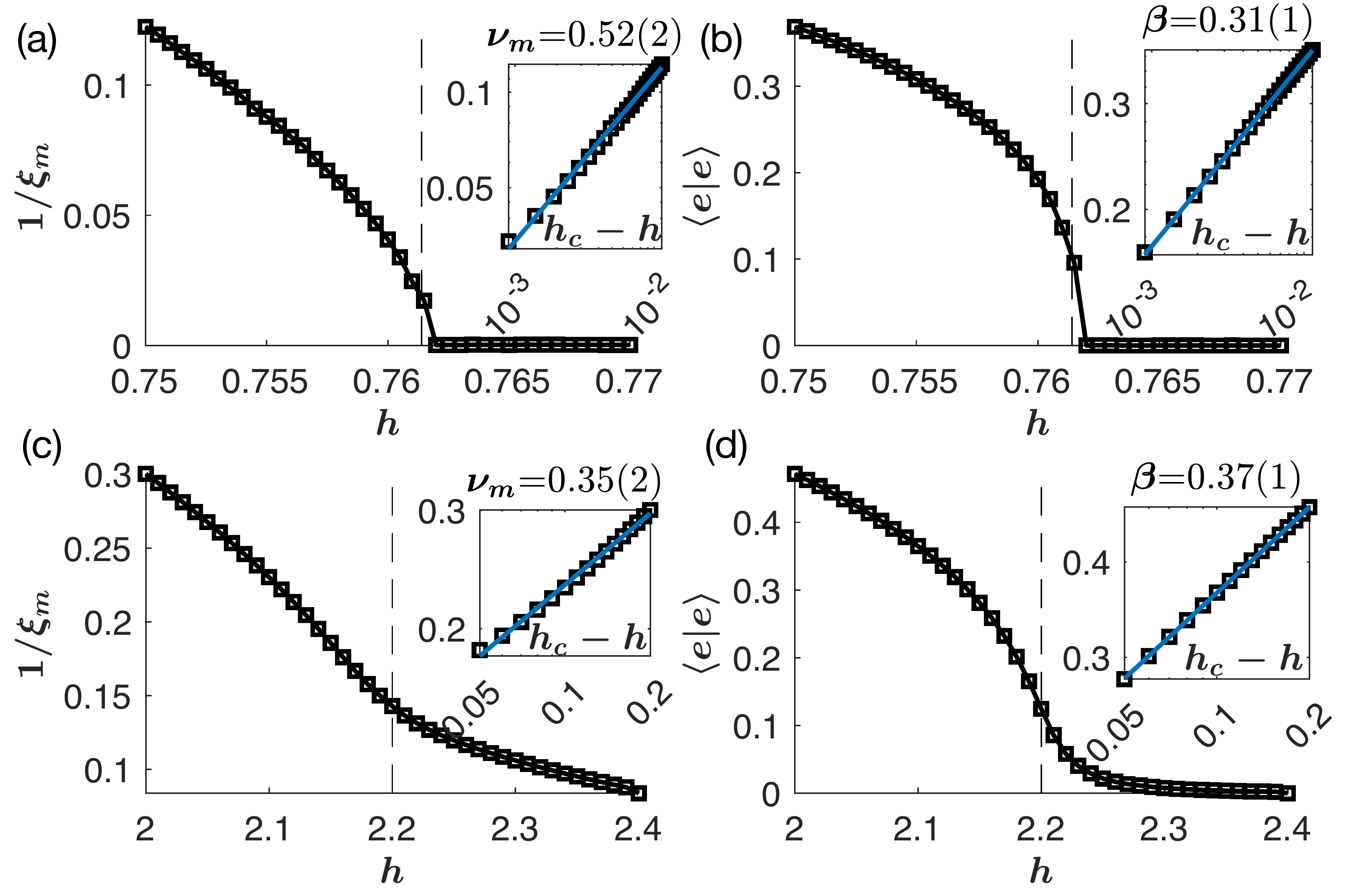}
    \caption{{\bf QPT of $m$-loop condensation in the  $\mathbb{Z}_N$ model} for (a)(b) $N=2$ and (c)(d) $N=5$.
    (a)(c) Inverse of the $m$-loop condensation length scale.
    (b)(d) Deconfined charge amplitude. 
    Insets fit the critical exponents $\xi_m \propto |h_c-h|^{-\nu_m}$ and $\langle e|e \rangle \propto |h_c-h|^\beta$.
    Data are computed by iPEPS of bond dimension $(D=3,\chi=72)$ for $N=2$ and $(D=2,\chi=80)$ for $N=5$.
    Dashed vertical line denotes the critical point $h_c\approx 0.7614$ for $N=2$ and $h_c\approx 2.2$ for $N=5$.
}
\label{fig:mloopcond}
    \end{figure}

\textit{m-loop condensation.--}
This QPT is captured by the classical vector gauge model. 
A classical Wilson loop around an arbitrary membrane $M$ excites an $m$-loop excitation above the ground state, which we denote as $| \prod_{p\in \partial M}m_p\rangle$.
Consequently, the condensate fraction of $m$-loops can be measured by its overlap with the ground state
\begin{equation}
    \langle \psi | \prod_{p\in \partial M}m_p \rangle =
    \langle \prod_{p\in \partial M} s_p \rangle 
    = \langle \prod_{\square\in M}    W_\square \rangle
    \equiv e^{-|M|/\xi_m^2} \,,
    \label{eq:mloopcond}
\end{equation}
where $|M|$ denotes the area of $M$, and $\xi_m$ defines the $m$-loop condensation length scale beyond which larger loops are orthogonal to the ground state. In the TN representation it is a membrane correlation written as a product of $W_\square$. 
An $X$ operator to the virtual variable violates the local Gauss law and creates an $e$-particle, equivalent to a semi-infinite 't Hooft string in the classical gauge theory, measuring the deconfined charge amplitude $\langle e|e\rangle $. 
Using Wegner's Ising-gauge duality~\cite{Wegner71duality}, $\langle e|e\rangle $ is mapped to the dual Ising order parameter,
and the critical point $h_c\approx 0.7614$ can be deduced from the known 3D Ising critical temperature $2/\ln\coth{h_c}\approx 4.5115$~\cite{Xiang12ising3dHOTRG}. 
In Fig.~\ref{fig:mloopcond}ab our iPEPS calculation shows that the loop condensation length scale $\xi_m$ is finite if $h<h_c$ and diverges for $h\geq h_c$; near the Ising* critical point it obeys a scaling law with exponent $\nu_m=0.52(2)$.
The critical exponent $\beta$ for the deconfined charge amplitude is also close to the Ising order parameter exponent $0.3295$~\cite{Xiang12ising3dHOTRG}.
Thus the divergence of the $m$-loop fluctuation length and the vanishing of the deconfined boson charge amplitude signal a continuous phase transition from the 2D TCs into the XC in the Ising* universality class~\cite{Trebst07toriccode}. 
    
\textit{Fracton confinement.--}
This QPT is described by the classical plaquette model. 
The model is invariant under planar subsystem Ising symmetries, which interpolate between the global and the local gauge symmetry~\cite{Wegner94PIM, Wegner98PIM, MooreXu04PIM, MooreXu05PIM} and are spontaneously broken across a first-order transition \cite{Johnston14PIM, Johnston17PIM} at $t_c  \simeq 0.66$.  
For an arbitrary membrane $M$, the probability amplitude for finding fractons at its corners (denoted by $\partial\partial M$) is measured by the wave function norm
\begin{equation}
    \norm{\ket{ \prod_{j\in \partial\partial M} f_j }}^2 
    =
    \langle \prod_{j\in \partial\partial M} \tau_j \rangle 
    =
    \langle \prod_{\square\in M} W'_\square \rangle 
    \equiv e^{-|M|/\xi_{f}^2} \,,
    \label{eq:fconf}
\end{equation}
where $f_j$ denotes the fracton, and $\xi_f$ defines the fracton confining length scale, beyond which the fracton amplitude decays exponentially. $1/\xi_f^2$ is analogous to the string tension in the quark confinement~\cite{Creutz80wilsonloop}.
Moreover, an $X$ operator to the virtual variable violates the magnetic Gauss law locally and creates a vector monopole excitation~\cite{supplemental}, whose expectation value measures the monopole condensate. 
An individual $Z$ operator in the TN evaluates the classical plaquette operator $W_\square'$, which corresponds to the probability of a fracton quadrupole around an elementary plaquette $\norm{\ket{ \prod_{j\in\square}f_j }}^2$, which is a composite particle freely mobile in all directions. 
\begin{figure}[b]
    \centering
    \includegraphics[width=\columnwidth]{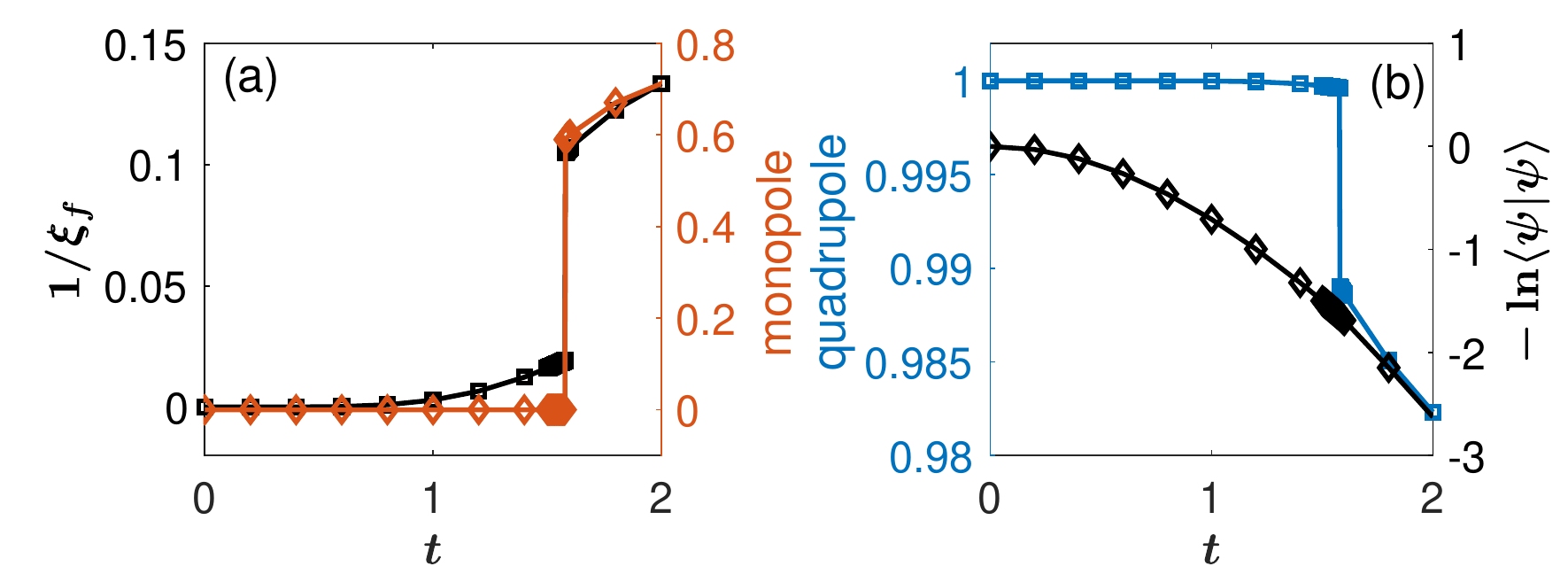}%
    \caption{{\bf Fracton confinement transition in the $\mathbb{Z}_N$ model} for $N=24$. $t_c\approx 1.5788(13)$.
    (a) Left axis: inverse fracton-confinement lengthscale; right axis: monopole condensate fraction. 
    (b) Left axis: fracton quadrupole amplitude; 
    right axis: logarithm of the wave function norm mapped to the classical free energy density. 
    Data are computed with iPEPS with bond dimension $D=2,\chi=24$. 
    }
    \label{fig:fractonconf}
\end{figure}
As shown in Fig.~\ref{fig:fractonconf} and the SM, our iPEPS calculations indeed confirm the first-order transition across which the fracton confining length jumps from a large value for $t<t_c$ to a finite value when $t>t_c$, accompanied by a jump of the monopole condensate $\langle\psi|\text{monopole}\rangle$ from approximately zero to a finite value. The fracton quadrupole amplitude and the second Rényi entropy coefficient also exhibit a jump from approximately 1 to a finite value, and the effective free energy density as a generating function shows a clearly visible kink
\footnote{We note that the nonzero $1/\xi_f$ at $t<t_c$ is due to the finite bond dimension effect of the iPEPS ansatz which cannot truly capture infinite correlation length, and $\xi_f$ shall grow larger with larger bond dimension (see SM).}.
For $t<t_c$ the wave function with isolated fracton excitations and the degenerate ground states on a torus~\cite{Fu16, Hermele17} are renormalizable and well defined. For $t>t_c$, the state with fractons separated at large distances has exponentially vanishing norm and is thus unrenormalizable, which means the fracton excitations as well as the topological degeneracy are gone in the thermodynamic limit -- a hallmark for the breakdown of topological order. 
While $\langle\psi|\psi\rangle$ is interpreted as a partition function, from $\norm{\ket{\prod_{j\in \partial\partial M} f_j }}^2\propto e^{-F(M)}$ one may define a dimensionless free energy $F(M)=|M|/\xi_f^2$ which captures the energetics of a set of fracton excitations lying at the corners of $M$. $F(M)$ 
reflects the underlying entanglement structure of the ground state wave function~\cite{Verstraete13toriccode, Verstraete15shadow, Schuch21entanglementorder} 
, reminiscent of the fact that the dimensionless entanglement Hamiltonian from a pure ground state can also capture the low energy behavior of a true physical boundary~\cite{Haldane08entspec,Cirac11entspec,Qi12entspec}.


\begin{figure}[t] 
   \centering
    \includegraphics[width=\columnwidth]{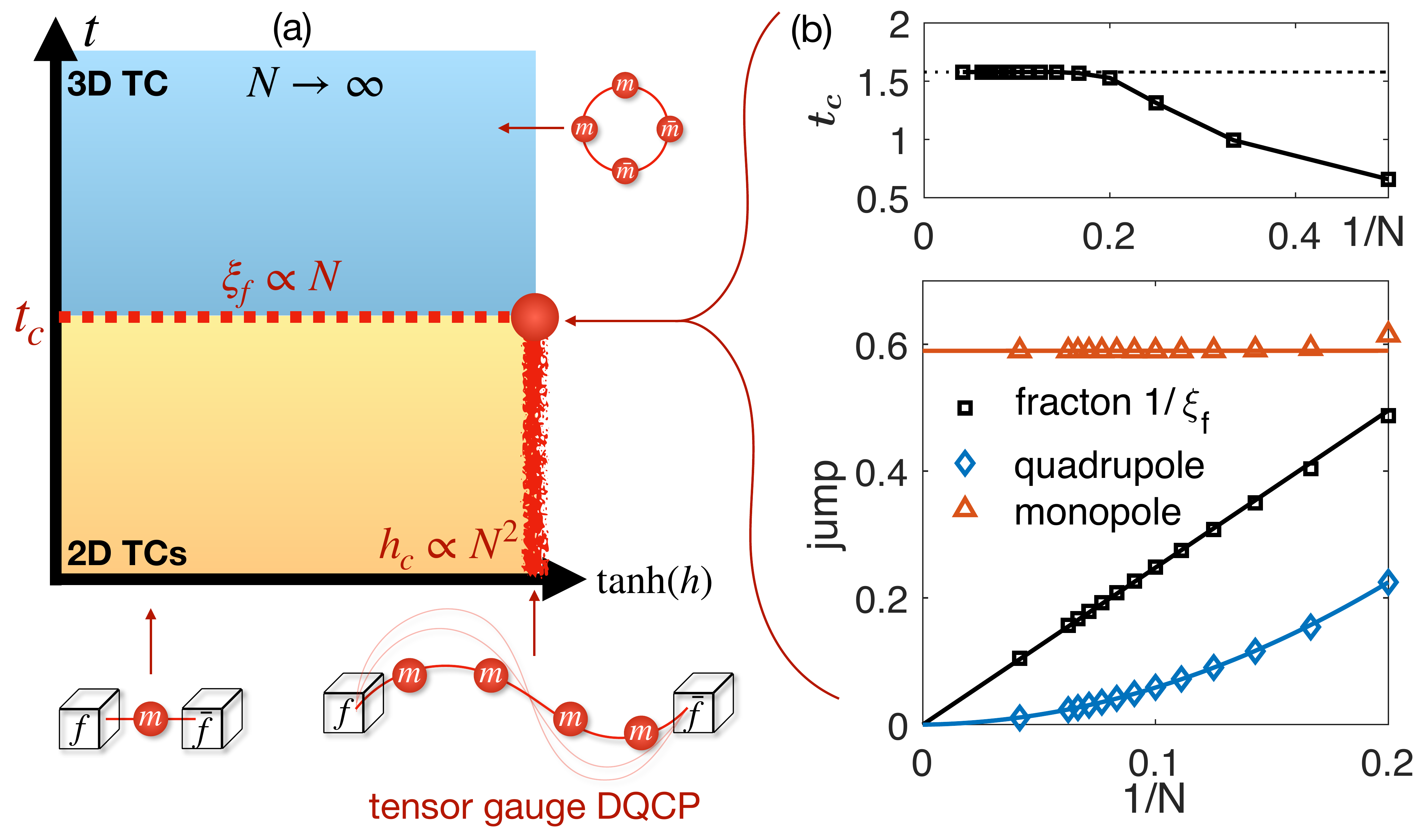} 
   \caption{{\bf Large-$N$ limit}. (a) Schematic phase diagram. 
   The XC fracton phase shrinks to a quantum critical line at $\tanh{(h_c)}=1$, on which deconfined fractons are glued by critical fluctuating $m$-strings. $h$ perturbation drives a partial confinement that binds fractons into fracton dipole, adiabatically connected to the $m$-particle in 2D TCs. $t>t_c$ drives complete confinement that leaves no deconfined fracton dipoles. 
   (b) $N$-scaling for the first-order QPTs of $\mathbb{Z}_N$ plaquette model: 
   $t_c$ converges to $t_c\approx1.5775(25)$. 
   The jump of the monopole condensate fraction remains finite and largely independent of $N$. 
   The fracton confining length $\xi_f$ for $t_c^+$ is proportional to $N$.
   The jump of fracton quadrupole amplitude vanishes like $1/N^2$.
   Data are computed with bond dimension $D=2$ for $N=2,\ldots,15,16,24$.
   }
   \label{fig:znplaqscaling}
\end{figure}

\textit{$\mathbb{Z}_N$ generalization.--}
All of the above can be generalized to the  
$\mathbb{Z}_N$ gauge group, which interpolates between $\mathbb{Z}_2$ and the compact $U(1)$ gauge group.
Importantly, the factorization \eqref{eq:factorization} still holds and gives rise to a $\mathbb{Z}_N$ lattice vector gauge model and a $\mathbb{Z}_N$ generalization for the plaquette Ising model with planar $\mathbb{Z}_N$ subsystem symmetries~\cite{supplemental}.
The $\mathbb{Z}_N$ vector gauge model \cite{Creutz80,Papa14} on a cubic lattice has been studied 
by using Kramers-Wannier duality to map it onto classical clock spin models exhibiting a single 
\footnote{
The clock anisotropy  is dangerously irrelevant at the 3D $XY$ critical point~\cite{Oshikawa00}, 
which indicates the absence of a stable intermediate Coulomb phase in the thermodynamic limit, and thus only a single transition. 
}
 transition depending on $N$: for $N\geq 5$ it is believed to undergo a continuous phase transition in the 3D $XY$ universality class~\cite{Creutz80,Oshikawa00,Papa14}. 
In our context, this implies that the $m$-loop condensation transition of our phase diagram in Fig.~\ref{fig:phasediagram} persists in the generic $\mathbb{Z}_N$ scenario (except for $N=3$ where the transition at $h_c$ becomes first order).
Our iPEPS calculation for the $\mathbb{Z}_5$ case in Fig.~\ref{fig:mloopcond}cd shows that the $m$-loop condensation length and the deconfined charge order parameter indeed approximately follow the conjectured scaling of the 3D XY universality class~\cite{Vicari013DXY}.
For the fracton confinement transition, on the other hand, our iPEPS calculations for the  plaquette clock model with finite $N=2,3,\ldots, 15,16, 24$ indicate first-order transitions with finite jumps similar to the $\mathbb{Z}_2$ scenario (see Fig.~\ref{fig:fractonconf} and SM), which however become notably weaker with increasing $N$.


\textit{$N\to\infty$ limit.--}
Of particular interest thus is the asymptotic limit in which $\mathbb{Z}_N$ approaches the compact $U(1)$ gauge group. 
For the vector gauge model, Monte Carlo simulations~\cite{Creutz80,Papa14} found that $h_c\propto N^2$, consistent with the absence of a deconfined $U(1)$ vector gauge phase in 3D~\cite{Polyakov77}. For the plaquette model, we however find that the critical point converges to a {\sl finite} value  $t_c\to 1.58$ (Fig.~\ref{fig:znplaqscaling}). The inverse fracton confining length $1/\xi_f$ decreases $\propto 1/N$ (for $t_c^+$), and the transition jump of the quadrupole amplitude vanishes $\propto 1/N^2$.  
These numerical observations indicate that the deconfined $\mathbb{Z}_{N\to\infty}$ XC fracton phase shrinks to a {\sl critical line} $(h_c\propto N^2, t\leq t_c)$ in the asymptotic limit. 

What is the nature of this critical line?
The asymptotic limit can be described by the unHiggsed $U(1)$ hollow tensor gauge theory studied in Refs.~\cite{XuWu08RVP, Chen18tensorhiggs,Seiberg21ZNfracton}, where the electric field tensor is purely off diagonal so as to enhance the dipole conservation to subsystem charge conservation as an organizing principle~\cite{supplemental}. Also, the classical plaquette model that our wave function is mapped onto, is equivalent to the classical field theory predicted for the Rokhsar-Kivelson point~\cite{XuWu08RVP}. 
Such a 3D quantum tensor gauge theory is known to be generally unstable against monopole proliferation ~\cite{XuWu08RVP}, which gaps out a deconfined Coulomb phase, analogous to the instability of a 2D quantum $U(1)$ vector gauge theory~\cite{Polyakov77}. 
Nevertheless, it does not rule out the possibility that the deconfined gauge theory can emerge at a critical point, dubbed \textit{deconfined quantum critical point} (DQCP)~\cite{Senthil04dqcp}. 
What we find here is a tensor gauge generalization of the vector gauge DQCP wave function~\cite{Zhu19ToricCode}, as the unHiggsed fracton phase shrinks to a  line of critical points decorated with critical fluctuating $m$-strings that glue pairs of fractons (Fig.~\ref{fig:znplaqscaling}). 
As a consequence, the $h$ perturbation is relevant (in the renormalization group sense) and confines two fractons into a fracton dipole. The fracton dipole is adiabatically connected to the deconfined $m$-particle in the 2D TCs phase, signalling a partial confinement from the fracton state. 
On the other hand, sufficiently strong $t>t_c$ leads to monopole proliferation and confines not only the fractons but also the fracton dipoles. 
Notice that the adjacent phases of our DQCP do not exhibit spontaneous symmetry breaking but rather distinct topological orders. 
It remains to be understood why the jump of  
the monopole condensation order across $t_c$ extrapolates to a {\sl finite} value, which is mapped to the nonlocal symmetry twist defect in the classical model.


\textit{Outlook.--}
The two phase transition lines cross at a peculiar multicritical point that can serve as a parent critical state, upon which any perturbations are relevant and flow to all four possible topologically distinct states. 
It is straightforward to generalize the isotropic deformation to anisotropic deformation, covering a richer variety of anisotropic subdimensional criticalities in Refs.~\cite{Kim17, Hermele21fracton, Williamson21SubdimensionalCriticality}. 
Our approach, by deforming an exact tensor network state and mapping to a tunable and computable classical statistical model, can be further extended to type II fracton orders as fractal condensates~\cite{Haah11, Yoshida13}, and twisted fracton topological order~\cite{Song19twistedfracton}. 
The wave function-deformed QPT of our study is particularly suitable for realization in programmable quantum simulators~\cite{Roushan21sycamoretoriccode, Lukin21toriccode, Lukin21quantumprocessor}, where the application of a local nonunitary circuit can directly deform the wave function. 
Our 3D tensor network wave function also serves as a natural variational ansatz for Hamiltonian deformations, which we leave to future work. 

\begin{acknowledgments}
\textit{Acknowledgments.--}
We acknowledge partial funding from the Deutsche Forschungsgemeinschaft (DFG, German Research Foundation) -- project Grant No. 277146847 -- through CRC network SFB/TRR 183 (projects A03, A04, B01).
P.Y. was supported by NSFC Grant  No.~12074438, Guangdong Basic and Applied Basic Research Foundation under Grant No.~2020B1515120100, and the Open Project of Guangdong Provincial Key Laboratory of Magnetoelectric Physics and Devices under Grant No.~2022B1212010008.
J.-Y. C. acknowledges support
from Sun Yat-sen University through a startup grant.
 The numerical simulations were performed on the CHEOPS cluster at the RRZK Cologne and on the JUWELS cluster at the Forschungszentrum Juelich using NLopt package~\cite{NLoptPackage}.
\end{acknowledgments}

\bibliography{FractonQPT}


\clearpage
\appendix
\tableofcontents


\section{Numerical phase diagrams of $\mathbb{Z}_N$ plaquette model}
\begin{figure}[ht]
    \centering
    \includegraphics[width=\columnwidth]{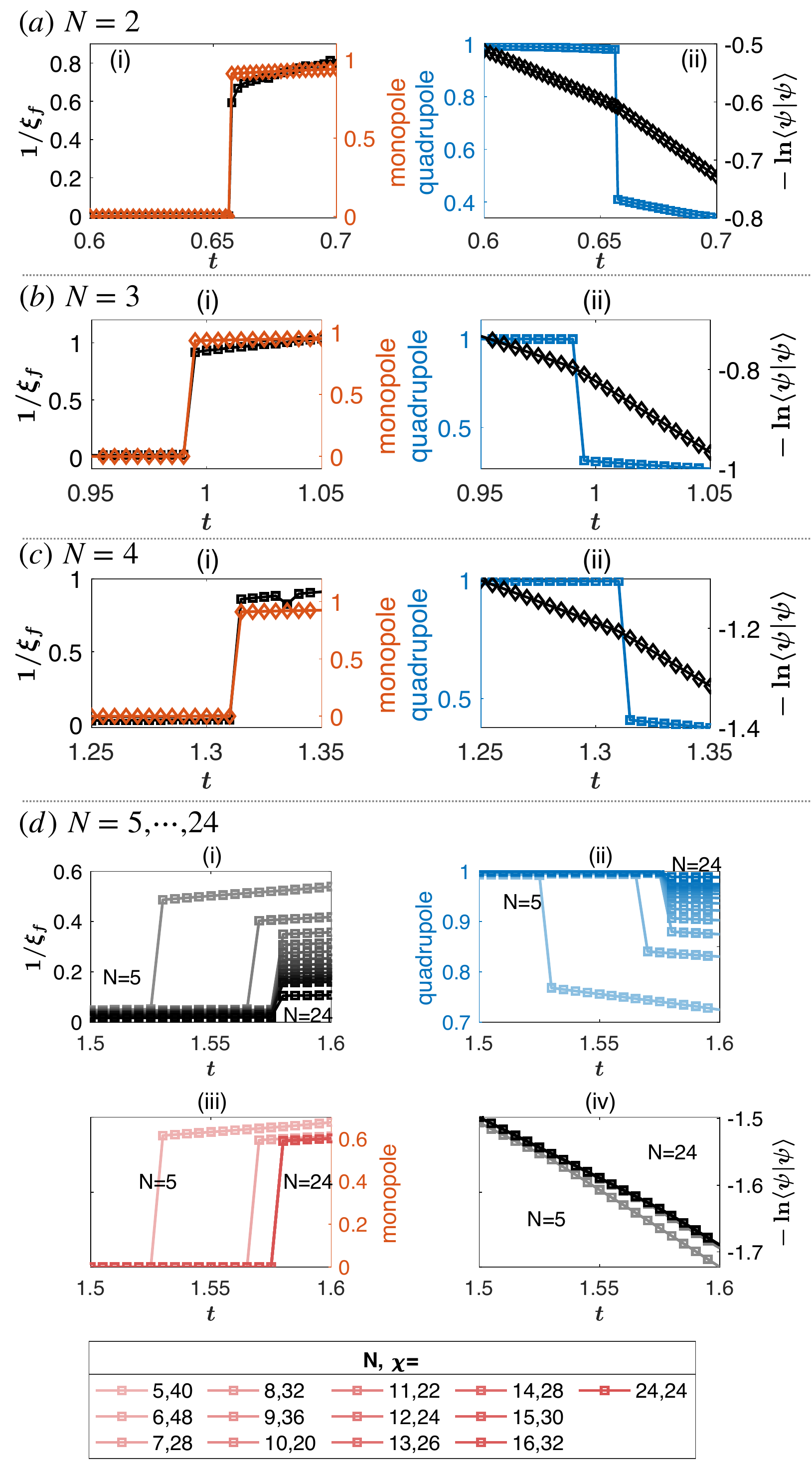}
    \caption{{\bf $\mathbb{Z}_N$ plaquette model} for $N=2,3,4,5,6,\ldots,15,16,24$, calculated by variational iPEPS of fixed bond dimension $D=2$, for which the boundary MPS dimension $\chi$ is also fixed to be integer multiples of $D^2$ or $D^2N$.
    }
    \label{fig:Z3to24plaq}
\end{figure}

Let us start by highlighting one of our key numerical results -- the morphing of the first-order fracton confinement transition captured by the $\mathbb{Z}_N$ plaquette model into a continuous phase transition as one approaches the $N \to \infty$ limit. 
In the main text, we had shown data for the two extreme cases of $N=2, 24$ in Fig.~3.
Here we want to fill out some of the intermediate values of by explicitly showing the numerical scans of the phase diagrams for $N=3,4,\ldots,15,16,24$ in Fig.~\ref{fig:Z3to24plaq} (taken with a resolution of $\Delta t=0.005$, somewhat coarser than in the main text).
The data indicates that all finite $N$ cases behave qualitatively similar to the $N=2$ case, with the decreasing jumps in the fracton quadrupole amplitude (and a softened kink in the free energy) indicating the qualitative trend to weaker first-order transitions for larger $N$. The quantitative fitting of this finite-$N$ data and its extrapolation $N \to \infty$ is shown in Fig.~4 of the main text, where the transition point $t_c^+$ is determined as given in Table \ref{tab:transition_points} below.
In performing the analysis of Fig.4 in main text, we take the data at the critical point $t_c^+$ for each $N\geq 7$ to fit the following scaling forms: 
the jump of monopole condensate fraction extrapolates to a constant $0.5897(5)$; 
the inverse fracton confining length scales as $1/\xi_f \propto 2.478(15)/N$; 
the jump of the fracton quadrupole amplitude is found to fit $0.04721(1266)/N +  5.380(116)/N^2$. 

\begin{table}[ht]
\centering
\caption{$\mathbb{Z}_N$ fracton confinement transition points.}
\label{tab:transition_points}
\begin{tabular}[t]{c|c|c|c|c|c|c}
\toprule
 $N$ & 2 & 3&4&5&6&$\geq 7$\\
\midrule
$t_c^+$ & 0.660(5) & 0.995(5) & 1.315(5) & 1.530(5) & 1.570(5) & 1.580(5)\\
\bottomrule
\end{tabular}
\end{table}

\section{Frustration-free Rokhsar-Kivelson Hamiltonian}
Here we prove for the $\mathbb{Z}_2$ case that the deformed wave function in the main manuscript can indeed be written as a ground state stabilized by a frustration free Rokhsar-Kivelson Hamiltonian
\begin{equation}
\begin{split}
 H_{\rm RK} =& - \sum_{+} \hat{A}_{+} - \sum_\square \hat{B}_{\square} + \sum_{\square}\hat{V}_{\square} + \sum_+\hat{V}'_{+},\\
\hat{A}_+ =& \prod_{l\in +}Z_{l}, \quad \hat{B}_{\square} = \prod_{l\in\square} X_{l},\\
 \hat{V}_{\square} =& e^{-h\sum_{l\in\square}\mu_l^z },\quad \hat{V}'_{+} = e^{-t\sum_{l\in +} \sigma_l^x },
    \end{split}
\end{equation}
 in the same spirit as in Ref.~\cite{Henley04RK}. Using the anti-commutation relation $\hat{B}_\square \hat{V}_\square(h)=\hat{V}_\square(-h)\hat{B}_\square$ one can show that
\begin{equation}
    (\hat{V}_\square(h) - \hat{B}_\square)^2 
    =(\hat{V}_\square(h) - \hat{B}_\square)(\hat{V}_\square(h) + \hat{V}_\square(-h)).
\end{equation}
Because $\hat{V}_\square(h) + \hat{V}_\square(-h)\geq0$ and $(\hat{V}_\square(h) - \hat{B}_\square)^2 \geq 0$, 
\begin{equation}
\hat{V}_\square(h) - \hat{B}_\square \geq 0,
\end{equation}
as a positive semi-definite operator. 
Likewise, for each inplane star vertex $+$, using $\hat{A}_+ \hat{V}_+'(t)= \hat{V}_+'(-t)\hat{A}_+$ one can show that 
\begin{equation}
    (\hat{V}_+'(t) - \hat{A}_+)^2 
    =(\hat{V}_+'(t) - \hat{A}_+)(\hat{V}_+'(t) + \hat{V}_+'(-t))\geq0,
\end{equation}
proving $\hat{V}_+'(t) - \hat{A}_+$ to be positive semi-definite operator. Therefore the Hamiltonian is a sum of positive semi-definite operators with positive semi-definite energy spectrum. Then we show that our deformed wave function is annihilated by this Hamiltonian i.e. having the lowest energy $0$. By using $|\psi_0\rangle = \hat{B}_\square|\psi_0\rangle = \hat{A}_+|\psi_0\rangle$, one can verify
\begin{equation}
\begin{split}
    \hat{B}_\square |\psi\rangle 
    &= \hat{B}_\square e^{\frac{1}{2} \sum_l t\sigma_l^x + h \mu_l^z}|\psi_0\rangle\\
    &= \hat{V}_\square(h) e^{\frac{1}{2} \sum_l t\sigma_l^x + h \mu_l^z}\hat{B}_\square|\psi_0\rangle
    = \hat{V}_\square(h) |\psi\rangle,\\
    \hat{A}_+ |\psi\rangle 
    &= \hat{A}_+ e^{\frac{1}{2} \sum_l t\sigma_l^x + h\mu_l^z} |\psi_0\rangle\\
    &= \hat{V}_+'(t) e^{\frac{1}{2} \sum_l t\sigma_l^x + h\mu_l^z} \hat{A}_+|\psi_0\rangle
    = V_+'(t) |\psi\rangle.
    \end{split}
\end{equation}
Therefore $\ket{\psi}$ is the ground state of this $\mathbb{Z}_2$ Rokhsar-Kivelson type Hamiltonian since $(V_+'-\hat{A}_+)\ket{\psi} = (\hat{V}_\square - \hat{B}_\square) \ket{\psi} = 0$.

More generally, a gapped PEPS is guaranteed to have a family of frustration-free parent Hamiltonians as a sum of quasi-local projectors that project onto the nullspace (kernel) of the linear map from the physical space to the virtual space i.e. excited states in the Hilbert space, and the uniqueness of the ground state is related to the injective property of this map while the topological degeneracy is related to the virtual entanglement symmetry~\cite{Cirac21rmp}.


\section{$\mathbb{Z}_N$ generalization of coupled toric code layers}

\begin{figure}[t]
    \centering
    \includegraphics[width=\columnwidth]{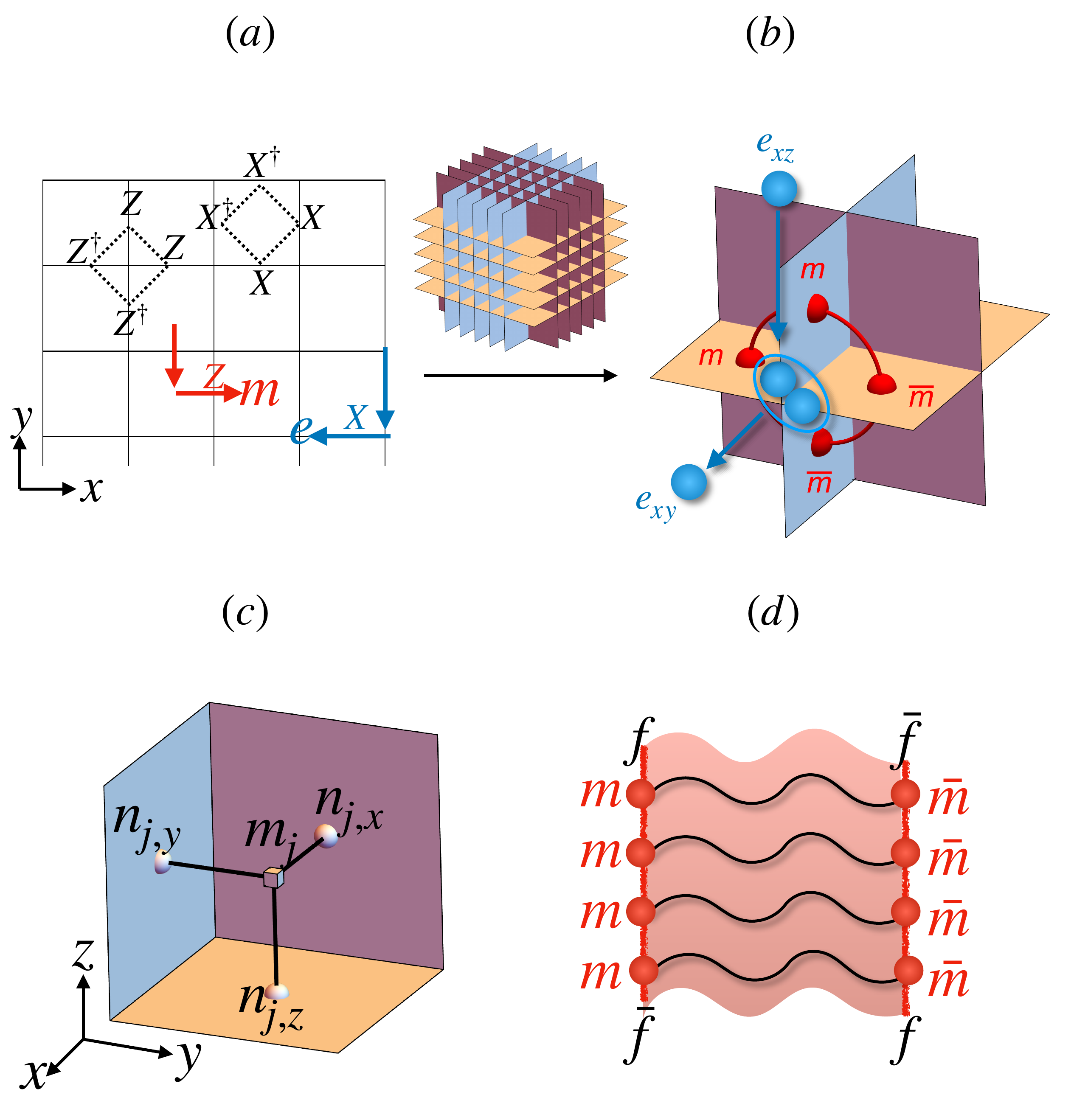}
    \caption{
    (a) $\mathbb{Z}_N$ 2D TC model. 
    (b) Coupled 2D TC layers.
    (c) Cube-vertex duality.
    (d) Fracton is the source of string of massive $m$ particles. Each $m$ particle is the source of a fluctuating string condensed in the vacuum of each 2D TC layer. A fluctuating membrane woven from the fluctuating strings hosts the fractons at its corners. The energy cost of the membrane does not scale with its size in the XC phase, but scales proportional to its side-perimeter in the 2D TCs phase, and proportional to its area in the trivial PM phase. }
    \label{fig:TClayers}
\end{figure}

For the $\mathbb{Z}_N$ group, the Pauli matrices are generalized to the clock (diagonal) matrix $Z$ and the shift (off-diagonal) matrix $X$, whose elements are 
\[
Z_{ij} = \omega^j\delta_{i,j}, \quad\quad X_{ij}=\delta_{i-1,j} \,,
\]
where $\omega=e^{i2\pi/N}$. They satisfy $ZX=\omega XZ$ and are generally not Hermitian matrices. 
Unlike the $\mathbb{Z}_2$ scenario, there can be different conventions for defining the 2D $\mathbb{Z}_N$ TC, up to certain sublattice basis transformations. We use the translationally invariant convention of $\hat{A}_+$ and $\hat{B}_\square$ as shown in Fig.~\ref{fig:TClayers}a.

The exactly solvable Hamiltonian
\begin{equation}
H_0=-\sum_+\hat{A}_+ -\sum_\square\hat{B}_\square + h.c.
\end{equation}
stabilizes the ground state with $\hat{A}_+\ket{\psi_0}=\hat{B}_\square\ket{\psi_0}=\ket{\psi_0}$. The $\mathbb{Z}_N$ $e(m)$ particle is defined by the eigenstate with $\hat{A}_+(\hat{B}_\square)\ket{e(m)}=\omega\ket{e(m)}$ such that $X(Z)$ moves the $e(m)$ particle following the orientation as shown by the blue(red) arrows in Fig.~\ref{fig:TClayers}a. Notice that it takes $N$ number of $e(m)$ particles to fuse into the vacuum, and $N-1$ number of $e(m)$ on the same vertex(plaquette) may also be denoted as $\bar{e}(\bar{m})$ meaning an anti-particle taking negative charge(flux). Also notice that the positive and negative charges cost the same energy. 

For the 3D construction, consider parallel stacks of 2D TC in $xy$ planes, rotate them to $yz$ planes and $zx$ planes, and stack all of them together forming a $3L$ intersecting layers of TC~\cite{Hermele17}, where $L$ is the linear system size. 
As shown in Fig.~\ref{fig:TClayers}b, the onsite Ising coupling $ZZ$ creates two pairs of $m(\bar{m})$ excitations at the adjacent four plaquettes forming an elementary magnetic flux loop in 3D, while the onsite Ising coupling $XX^\dag$ hops a charge pair bound state following the blue arrow. In the following we denote the magnetic flux loop excitation composed of a loop of $m(\bar{m})$ particles as $m$-loop. 
The $m$-loop excitation is one of the elementary excitations in 3D TC, which does not allow open magnetic string excitation due to the magnetic Gauss law. One should not confuse the 3D $m$-loop excitation as a collection of $m$ particles that cost energy, with the string pulled by an individual $m$ particle that is condensed in the ground state in a single layer of 2D TC. But rather, the energetic $m$-loop excitation should be viewed as the boundary of a fluctuating membrane in 3D, just as $m$ particle is the source or drink of the fluctuating string in 2D. And the fluctuating membrane is in fact woven from the 2D fluctuating strings in this coupled-layer-construction, as shown schematically in Fig.~\ref{fig:TClayers}d. 
The $e_{xy}\bar{e}_{xz}$ paired composite particle is a monopole bound to the $x$ axis (Fig.~\ref{fig:TClayers}b), which is sometimes called a lineon in literature, since by definition it is pinned to the line of the two intersecting planes and unable to move away from that despite any perturbation, which is an apparent geometrical fact that two Euclidean planes intersect at not more than just one line. 

In this way the inplane 2D particles $e$ and $m$, also called planons, are reorganized into the electric charge and magnetic flux loop in the 3D TC subsystem (a vector gauge system), and the fractonic charge and magnetic monopole in the 3D XC subsystem (a tensor gauge system with only off-diagonal electric field tensor at plaquette center), see Table.~\ref{tab:ex2d3d}. 
However, before anyon condensation, each $e$ and $m$ particle costs energy, which means two fractons joined by an $m$ string cost energy linearly proportional to their distance, and are thus confined into a fracton dipole. 
Notice that the fracton dipole as a short open $m$ string segment is deconfined in 2D TCs but confined in the trivial phase. 
A bosonic charge moving from the $xy$ plane to the $xz$ plane inevitably leaves behind a monopole $e_{xy}\bar{e}_{xz}$ which costs energy. 
Therefore the stacked 2D TCs, despite the basis transformation, should be contrasted from the so-called hybrid fracton orders in Ref.~\cite{Vijay21hybridfracton}. 

\begin{table}[t]
\centering
\caption{Elementary electric and magnetic excitations in 3D TC and XC subsystems using $e$, $m$ particles in 2D TC layer as building blocks. }
\label{tab:ex2d3d}
\begin{tabular}[t]{c  |  c  |  c}
\toprule
 excitation & TC & XC\\
\midrule
electric & $e$-charge & $m$-string-source (fracton)\\
magnetic & $m$-loop & $e\bar{e}$-pair (lineon)\\
\bottomrule
\end{tabular}
\end{table}

Fig.~\ref{fig:TClayers}c shows the cube-vertex duality for the 3D cubic lattice, which swaps cubes to vertices, and links to plaquettes. The colored planes denote the original cubic lattice, while the black skeleton shows its dual counterpart. The quantum wave function for the 3D TC(XC) can be rewritten as a dual quantum paramagnet where the classical variables $\{n\}(\{m\})$ reside on the links(sites) of the dual cubic lattice while the physical variables reside on the links of the original lattice. 

The wave function of the stacked $\mathbb{Z}_N$ TC at its solvable fixed point is generalized to 
\begin{equation}
    |\psi_0\rangle
    = \sum_{\{n\}} |Z_{\partial p\cap \partial q}=\omega^{n_p - n_q}\rangle,
    \label{eq:Psi0}
\end{equation}
where $\{n=0,1,\ldots,N-1\}$ is the ensemble of virtual classical variables. 
In this abbreviated equation, since $Z\neq Z^\dag$ for $N>2$, one should particularly take care of the orientation from $p$ to $q$ in order to fulfill the star stabilizer $\hat{A}_+\ket{\psi_0}=\ket{\psi_0}$. For example, in the $xy$ plane of Fig.~\ref{fig:TClayers}a, $p=q-x$ or $p=q+y$. 
For the other planes, one cyclically permutes $(x,y)$ to $(y,z)$ and $(z,x)$. 

Again we can perform a basis transform and relabel the $\mathbb{Z}_N$ clock spin operators into
\begin{equation}
\begin{split}
    \mu_l^z\equiv Z_{l_1}Z_{l_2},\quad \mu_{l}^x\equiv X_{l_2},\quad 
    \sigma_{l}^x\equiv X_{l_1}X_{l_2}^\dag,\quad \sigma_{l}^z\equiv Z_{l_1}.
\end{split}
\label{eq:ZNCNOT}
\end{equation}
Here $l_1,l_2$ are defined respecting the cyclic permutation around the $(1,1,1)$ direction:
For example, take $l_1$ as the link originating from the $xz(yx)(zx)$-layered TC while $l_2$ from the $xy(yz)(zx)$-layered TC, although both of them point toward the $x(y)(z)$ direction, as follows:
\begin{equation}
\includegraphics[width=0.3\columnwidth]{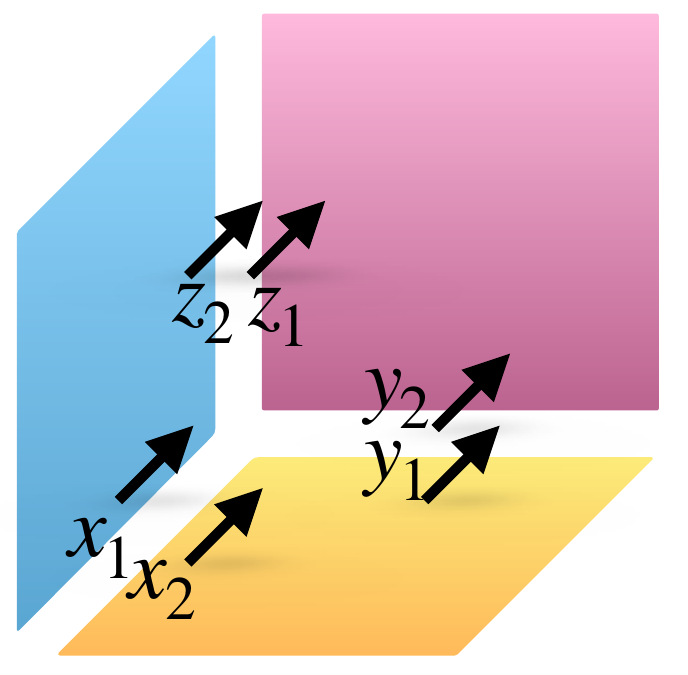}.
\end{equation}
In the ground state $\ket{\psi_0}$ the $\mathbb{Z}_N$ Gauss laws for vector gauge and tensor gauge theory are satisfied
\begin{equation}
\includegraphics[width=.666\columnwidth]{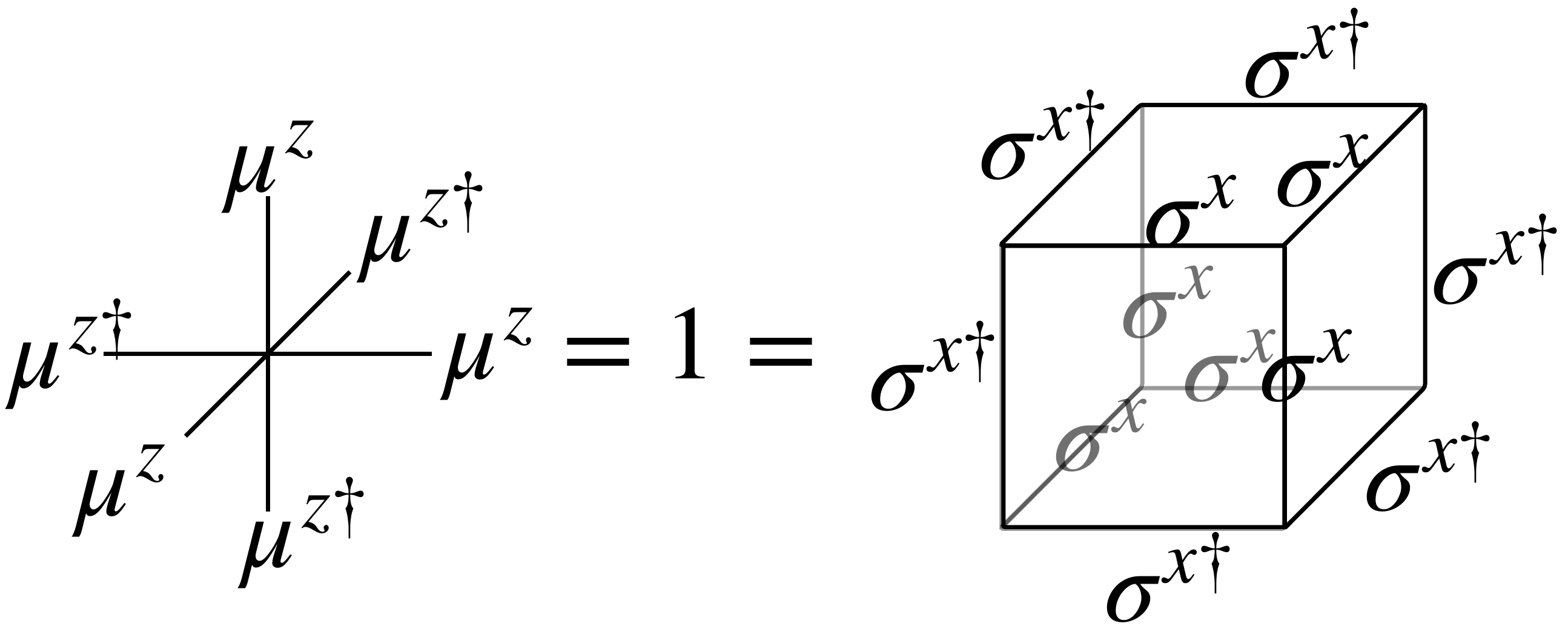}.
\end{equation}
Nevertheless, if tracing out either subsystem one can get a mixed state with proliferated vector-gauge magnetic flux loops, or tensor-gauge magnetic monopoles because
\begin{equation}
\bra{\psi_0} \prod_{l\in\square} \mu_l^{x(\dag)} \ket{\psi_0} =0=\bra{\psi_0} \prod_{l\in+} \sigma_l^{z(\dag)} \ket{\psi_0} .
\end{equation}
Our target is the deformed wave function as follows
\begin{equation}
    |\psi(t,h)\rangle \equiv e^{\frac{1}{2}\sum_l h\mu_{l}^z + t\sigma_{l}^x}|\psi_0\rangle \,.
\end{equation}


\section{Quantum classical mapping}
In this section, we will detail how to map the general $\mathbb{Z}_N$ coupled 2D TCs wave functions onto respective classical models. The key result is that the classical counterpart is always exactly factorized into the classical gauge and classical plaquette models. It is most straightforward to work in the vertex representation based on the $Z$-basis: $\ket{\psi_0}$ then is a superposition over all $\{Z\}$ configurations that satisfy the inplane vertex rules $\hat{A}_+=1$
\begin{equation}
\ket{\psi_0} = \sum_{\{Z\}} \prod_+ \frac{1+\hat{A}_+}{2}\ket{\{Z\}}.
\end{equation} 
For every unit-cell there are three independent $\mathbb{Z}_N$ constraints in total. 
Upon the basis transformation into $\{\mu,\sigma\}$, 
\begin{equation}
\includegraphics[width=\columnwidth]{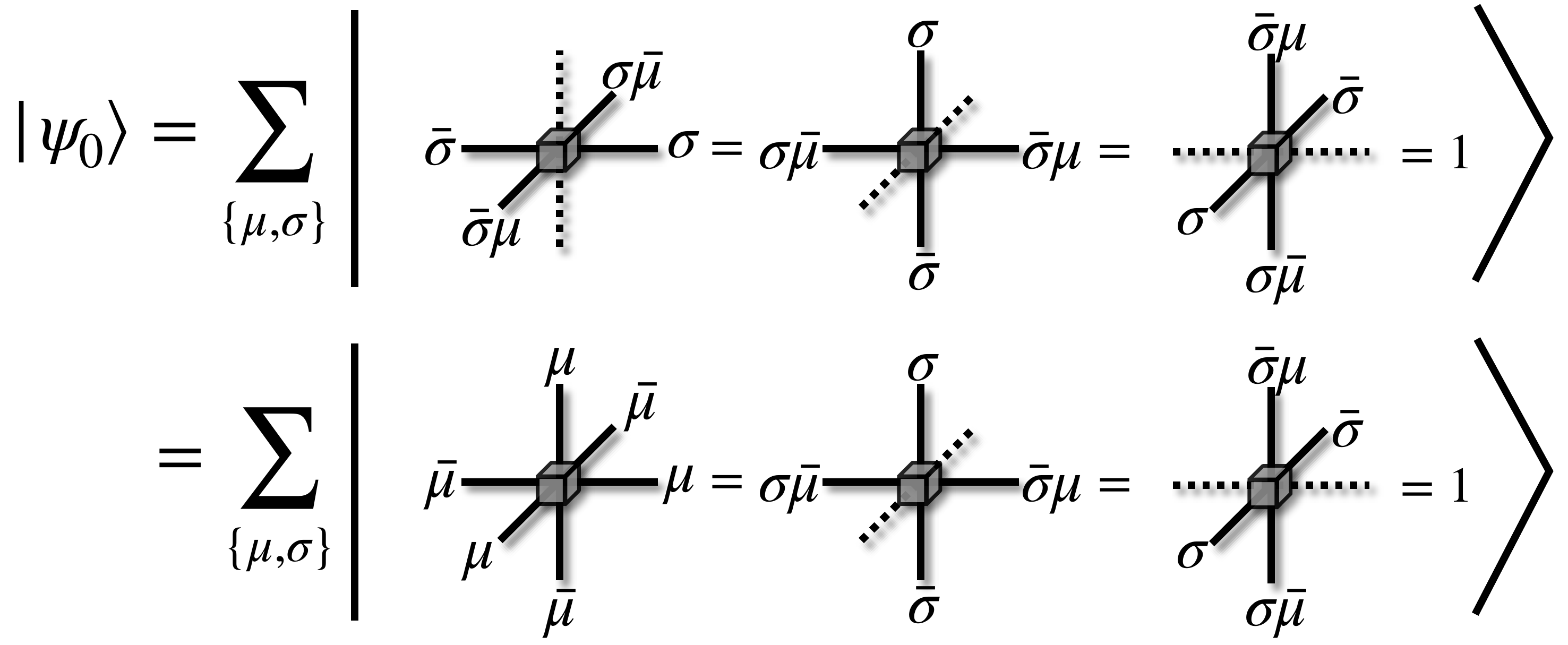},
\end{equation}
which becomes an entangled quantum vertex state. Physically speaking, the TC electric string turning point in $\{\mu\}$ subsystem is entangled with the XC monopole in $\{\sigma\}$ subsystem. 

Our objective is the partition function
\[
\bra{\psi_0} e^{\frac{1}{2}\sum_l (t\sigma_l^x + h\mu_l^z+h.c.)} \ket{\psi_0}
\]
obtained by contracting two hyper-layers of TN state for $\bra{\psi}$ and $\ket{\psi}$ joined by a set of local nonunitary gates $e^{(t\sigma^x + h\mu^z + h.c.)/2}$:
\begin{equation}
\includegraphics[width=.8\columnwidth]{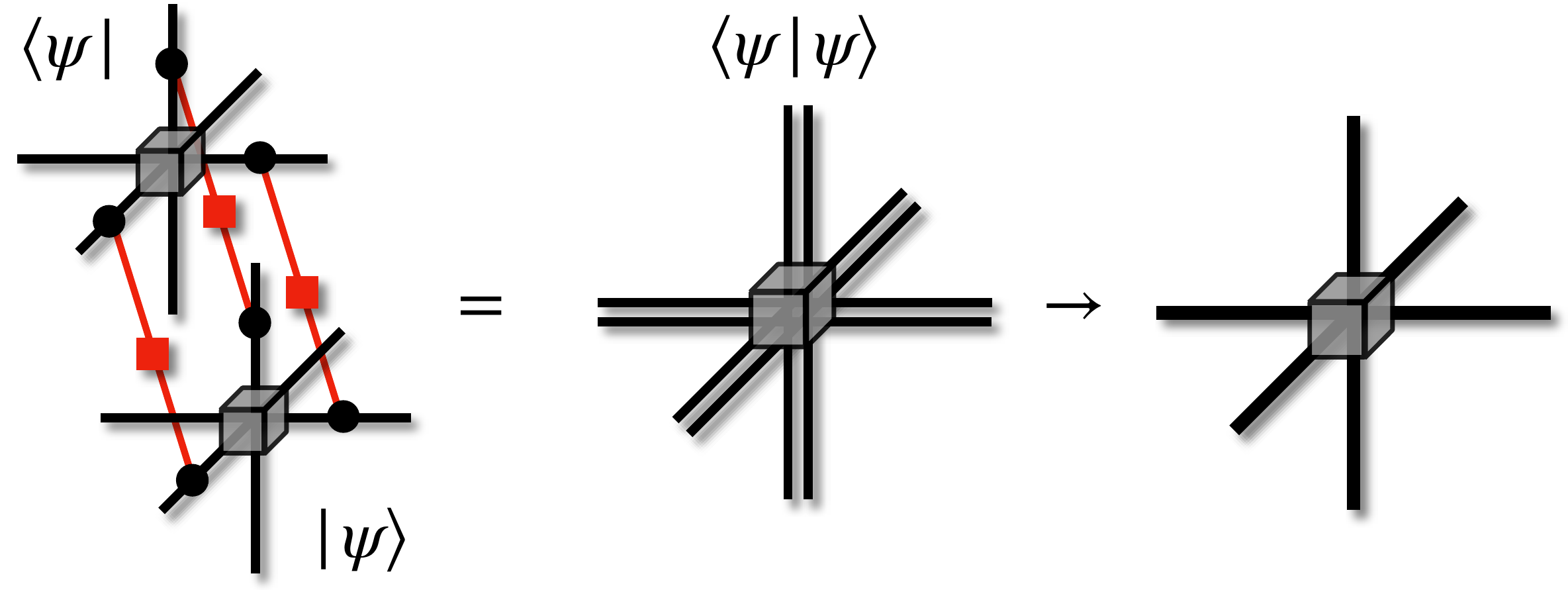},
\label{eq:brakettensor}
\end{equation}
where the red box denotes the local physical gate, and after contraction the double virtual indices are grouped into a thicker virtual index of dimension $N^2$, which can be further compressed into dimension $N$ as follows:
Firstly, $e^{h\mu^z/2+h.c.}$ is a local diagonal gate and seals $\{\mu\}$ in the ket layer and the bra layer to be identical and follow the Gauss law constraint:
\begin{equation}
\includegraphics[width=.25\columnwidth]{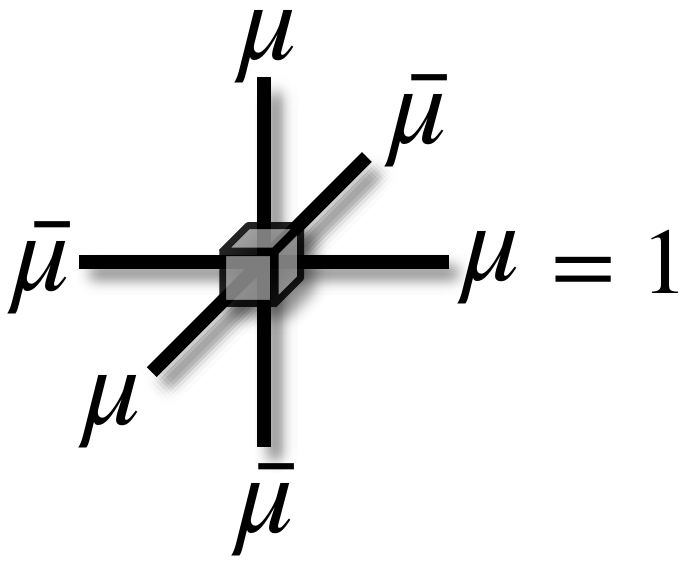}.
\label{eq:vertexrule1}
\end{equation}
Secondly, $e^{t\sigma^x/2+h.c.}$ as a rotation operator can connect $\{\sigma\}$ in $\ket{\psi_0}$ to an independent set $\{\sigma'\}$ in $\bra{\psi_0}$, where the offset $\{\bar{\sigma}'\sigma \equiv \omega^m\}$ is weighted by its Fourier coefficient
\begin{equation}
\begin{split}
 &e^{\frac{t}{2}(X + X^\dag)}  
    =\sum_{m=0}^{N-1} p_m^2(t) X^m,\\
    &p_m\equiv \sqrt {\frac{1}{N}\sum_{k} e^{t \cos{\frac{2\pi k}{N}}}\omega^{k m}},
    \end{split}
\end{equation}
where $X^m$ shifts the clock variable by $m$ units. 
Thirdly, observe that the vertex constraints in ket layer express the entanglement between $\{\mu\}$ and $\{\sigma\}$, and the same vertex constraints in bra layer express that between $\{\mu\}$ and $\{\sigma'\}$. Therefore the offset distribution $\{\bar{\sigma'}\sigma\}$ is free from being entangled with $\{\mu\}$ but is constrained to follow the latter two vertex constraints: 
\begin{equation}
\includegraphics[width=.5\columnwidth]{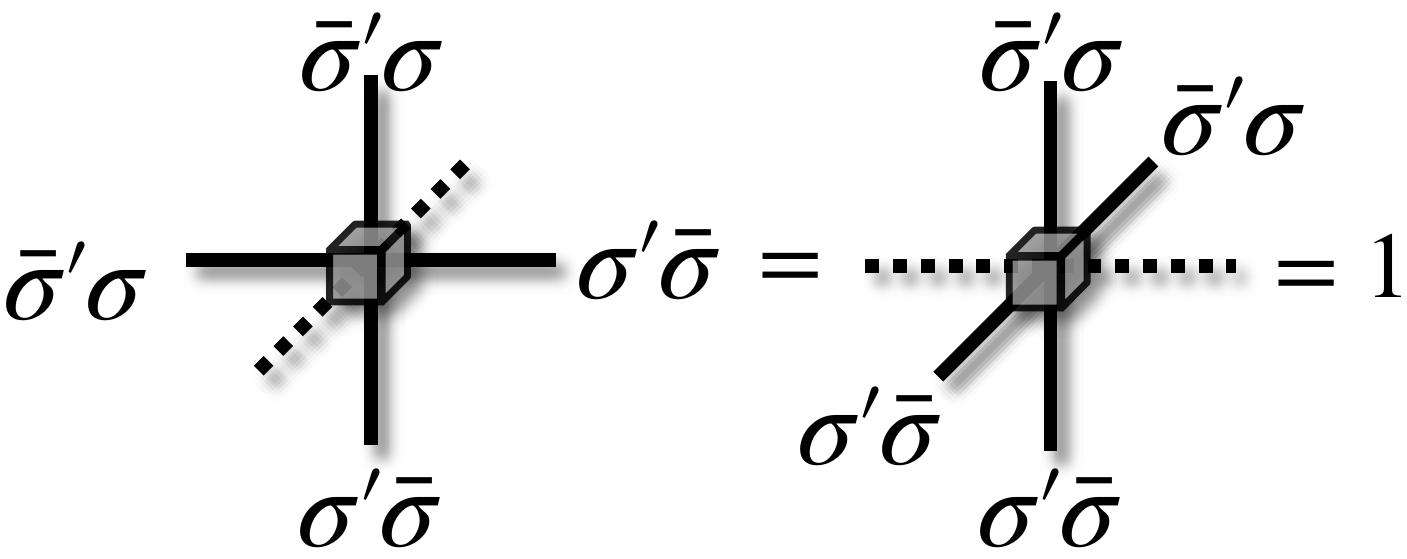}.
\label{eq:vertexrule2}
\end{equation}
We can further compress the variable $\bar{\sigma}'\sigma \equiv \tilde{\sigma}$, and perform a sublattice basis transformation that doubles the unit-cell to make the vertex rules for both $\mu$ and $\tilde{\sigma}$ mirror reflection symmetric. Finally, the partition function is factorized into a weighted average over all allowed vertex configurations that satisfy the vertex rules:
\begin{equation}
\includegraphics[width=.9\columnwidth]{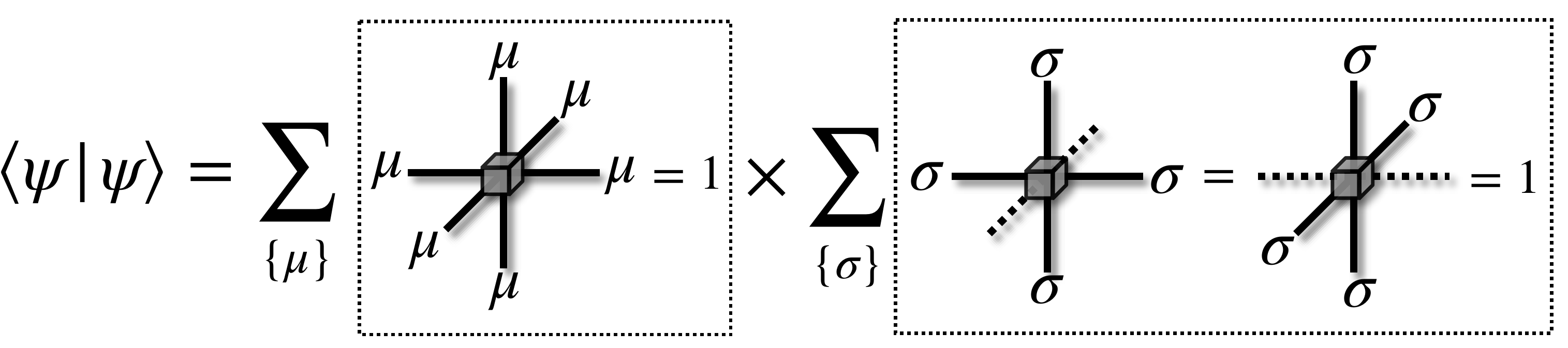},
\end{equation}
where each $\mu$ leg is weighted by $e^{h \text{Re}(\mu)/2}$ and each $\tilde{\sigma}\equiv \omega^n$ leg is weighted by $p_n$. 
Such classical vertex models are readily expressed as classical TNs, with each vertex being the local rank-6 tensors 
\begin{equation}
\begin{split}
&\hat{T}_g= \delta_{\sum_l n_l ,0} \times e^{\frac{h}{2}\sum_{l}\cos(\frac{2\pi}{N} n_l)},\\
&\hat{T}_p =
 \delta_{m_w+m_e+m_u+m_d,0}\delta_{m_n+m_s+m_u+m_d,0}\times \prod_{l} p_{m_l}.
\end{split}
\label{eq:ZNtensor}
\end{equation}
By the cube-vertex duality $l\leftrightarrow \square$, each leg variable can be represented by a dual plaquette variable $\mu_l\equiv W_\square$, $\tilde{\sigma}_l\equiv W'_\square$ to automatically satisfy the vertex rules \eqref{eq:vertexrule1} and \eqref{eq:vertexrule2} :
\begin{equation}
\includegraphics[width=\columnwidth]{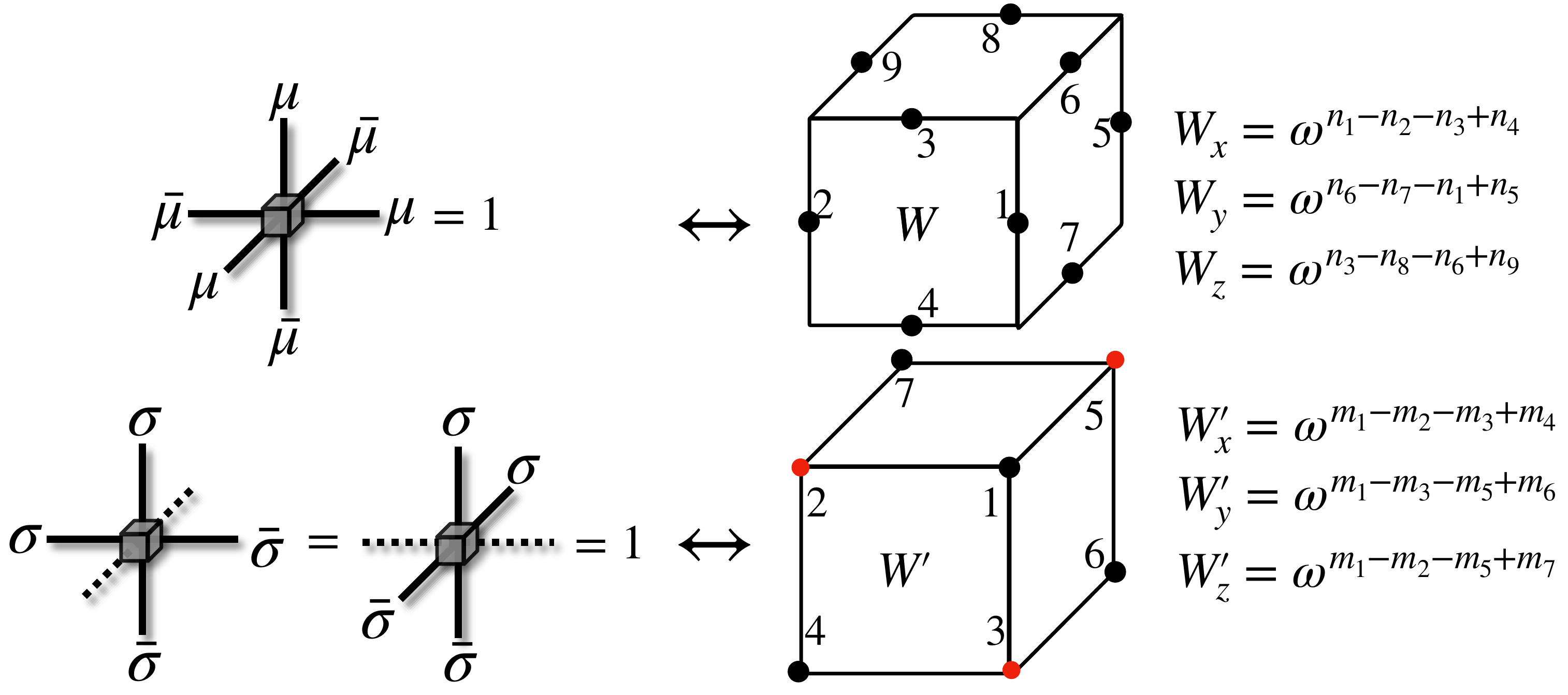},
\end{equation}
where $W$($W'$) is the $\mathbb{Z}_N$ plaquette interaction terms over the links(sites).
In this way $\langle \psi|\psi \rangle$ is mapped to a combination of classical models
\begin{equation}
\begin{split}   
&\epsilon_g = -\frac{h}{2}\sum_{\square} W_\square+h.c.,\quad\\
  & \epsilon_p = -\sum_{\square} \ln \left(\frac{1}{N}\sum_{k=0}^{N-1} e^{t \cos{\frac{2\pi k}{N}}} W_\square^{\prime k}\right) \,,
    \end{split}
    \label{eq:ZNpsithClassicModel}
\end{equation}
where $\epsilon_g$ is the standard $\mathbb{Z}_N$ lattice vector gauge model~\cite{Creutz80,Papa14}, while $\epsilon_p$ is a $\mathbb{Z}_N$ generalization for the plaquette Ising model with planar $\mathbb{Z}_N$ subsystem symmetries. 

Now we discuss the physical meaning of observables in the vertex TN and the classical models, explaining the dictionary table in the main text. Firstly, as seen from Eq.~\eqref{eq:brakettensor}, a diagonal operator $\mu$ acted on a vertex leg can be lifted to the physical spin $\mu^z$, which creates the $m$-loop excitation surrounding the leg. A product of such operators forming a membrane creates a large $m$-loop at the membrane boundary $\partial M$. In the classical gauge model it is equivalent to the Wilson loop observable because $\prod_{l\in M} \mu=\prod_{\square\in M} W_\square$.
Secondly, a membrane of diagonal operators $\tilde{\sigma}=\bar{\sigma}'\sigma$ inserted to the vertex leg is equivalent to inserting a membrane of $\sigma^z$ operators into both the ket $\ket{\psi}$ and the bra $\bra{\psi}$ in Eq.~\eqref{eq:brakettensor}, which creates four fractons at the membrane corners. An individual $\tilde{\sigma}$ operator corresponds to a tightly bound fracton quadrupole. In the classical plaquette model, the membrane corresponds to the four point corner correlation function, and each classical spin operator at each corner represents a fracton charge. 
These two diagonal membrane operators are schematically shown below:
\begin{equation}
\includegraphics[width=\columnwidth]{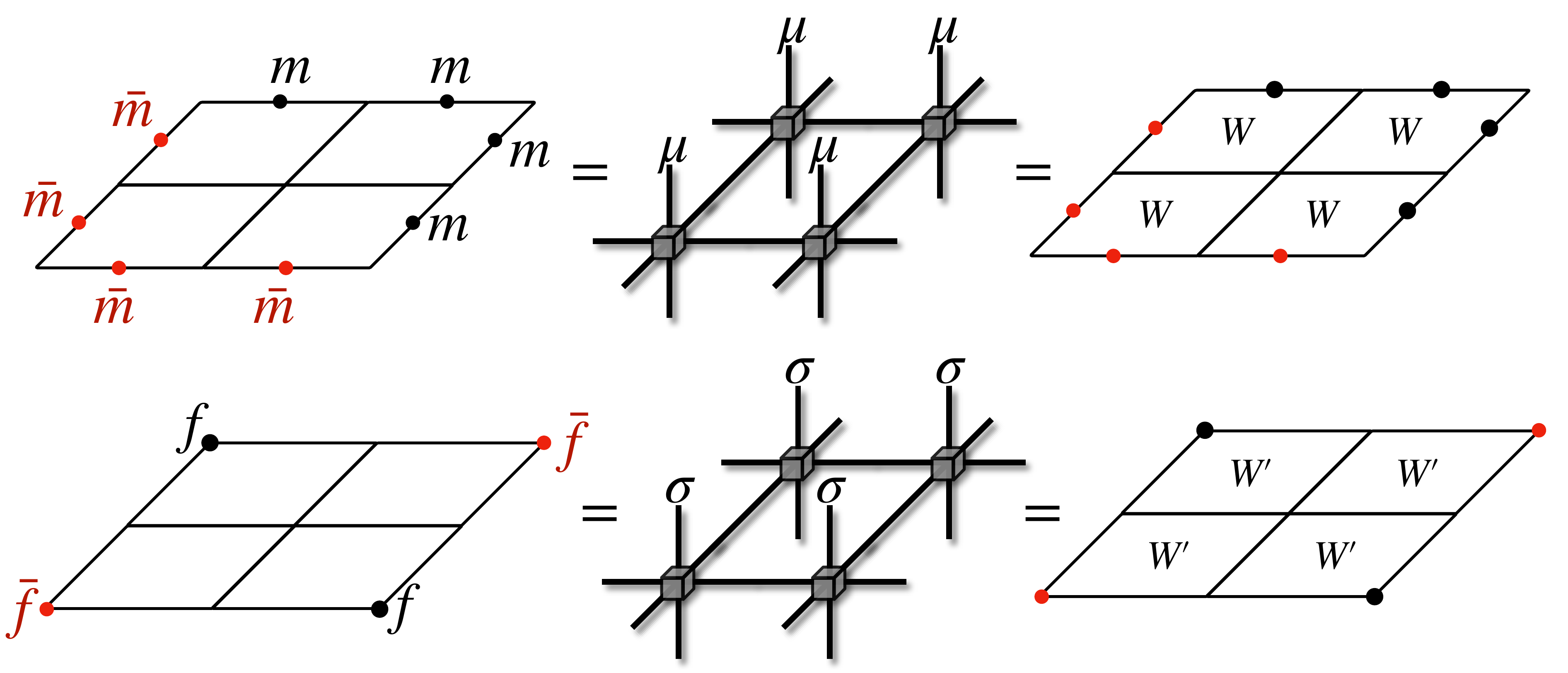}.
\end{equation}
Thirdly, an off-diagonal matrix $X$ inserted to a virtual leg can be absorbed to one adjacent vertex, which violates the vertex rules. The local violation of rule \eqref{eq:vertexrule1} corresponds to a boson charge defect, while the local violation of rule \eqref{eq:vertexrule2} corresponds to tunneling of a (lineon) magnetic monopole.


\section{X cube model and its tensor network wave functions}
\begin{figure*}[ht]
    \centering
    \includegraphics[width=\textwidth]{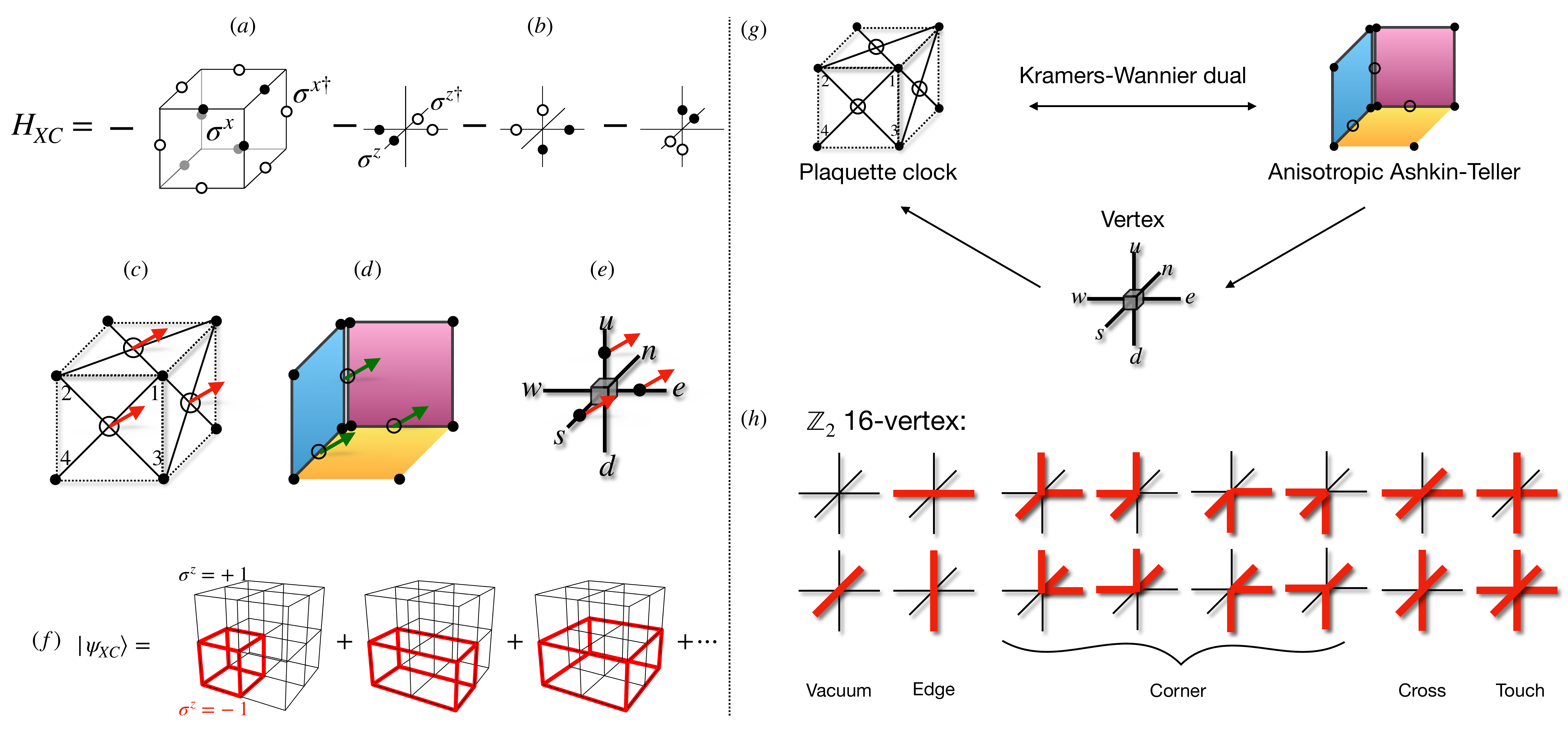}
    \caption{
   {\bf $\mathbb{Z}_N$ X cube Hamiltonian, its tensor network ground-state wave functions and corresponding classical models.}
   (a) the cube stabilizer;
   (b) Three inplane star stabilizers.
   (c) wave function as a dual plaquette clock paramagnet in $\sigma^z$ basis, generated by the rank-12 tensors of bond dimension $N$ at site joined by rank-5 tensors on plaquette centers.
   (d) wave function as a dual anisotropic Ashkin-Teller paramagnet in $\sigma^x$ basis, generated by rank-6 tensors of bond dimension $N$ in a simple cubic lattice.
   (e) wave function as a quantum vertex model in $\sigma^z$ basis, generated by the rank-6 tensor of bond dimension $N$.
   Black dots are for dummy virtual classical variables. 
      (f) The wave-function is a cuboid condensate as superposition of $\sigma^z=-1$ cuboids configurations.
      (g) The corresponding equivalent classical models where physical variables have been contracted.
      (h) For the simplest $\mathbb{Z}_2$ case, we show all the 16 classical vertex configurations as required by the star stabilizers in (a). The thin black link takes value $m=0$ while the bold red link takes $m=1$. The composition of the vertices leads to the cuboid condensate in (f). }
    \label{fig:XCtensor}
\end{figure*}
In this section we specifically discuss the pure X cube ground state wave-function, which has several equivalent representations and classical model counterparts, summarized in  Fig.~\ref{fig:XCtensor}. 
To start, the pure $\mathbb{Z}_N$  X cube model contains  three inplane star stabilizers and a cube stabilizer (Fig.~\ref{fig:XCtensor}ab). 
There are three equivalent TN representations for the its ground state wave function in Fig.~\ref{fig:XCtensor}cde, explained in the following:

\begin{itemize}[leftmargin=*]
\item {\it Representation}-1: dual plaquette clock paramagnet. $\sigma^z=\omega^{m_1-m_2-m_3+m_4}$ expresses $\sigma^z$ as the domain corner degree of freedom for the classical plaquette clock model.
In this way each classical configuration automatically favors the star stabilizers, and the superposition over classical configurations is enforced by the cube stabilizer
\begin{equation}
\begin{split}
\ket{\psi_0} =&\sum_{\{m\}}\prod_\square \ket{\sigma_\square^z=\omega^{m_1-m_2-m_3+m_4}}.
\end{split}
\end{equation}
As shown in Fig.~\ref{fig:XCtensor}c, it is readily generated by rank-12 tensors of bond dimension $N$ at sites, joined by rank-5 tensors at plaquette centers. However, this network does not have a simple cubic lattice geometry. 

\item {\it Representation}-2: dual anisotropic Ashkin-Teller paramagnet. There are two virtual classical $\mathbb{Z}_N$ spins at each site, as shown in Fig.~\ref{fig:XCtensor}d. By expressing the physical variable $\sigma^x$ as the domain wall of the classical spins, the cube stabilzer can automatically be favored in each classical configuration. The star stabilizers then superpose all the classical configurations
\begin{equation}
\begin{split}
\ket{\psi_0} =\sum_{\{m,n\}} 
&\ket{\sigma_{j,x}^x=\omega^{m_{j+x}-m_{j}}} 
\otimes\ket{\sigma_{j,y}^x=\omega^{n_{j+y}-n_{j}}}\\
&\otimes\ket{\sigma_{j,z}^x=\omega^{m_j+n_j-m_{j+z}-n_{j+z}}}.
\end{split}
\end{equation}
In this way it can be generated by a rank-6 tensor with bond dimension $N$. 
This TN, however, is not mirror reflection symmetric. 

\item {\it Representation}-3: quantum vertex model in $\sigma^z$ basis, shown in Fig.~\ref{fig:XCtensor}e. The virtual variables are taken to be identical to the physical $\mu^z=\omega^m$ where $m=0,\ldots,N-1$, which are subject to a vertex rule declared by the star stabilizers
\begin{equation}
\begin{split}
\ket{\psi_0} =&\text{tTr}\prod_j 
\delta_{m_w-m_e+m_u-m_d,0}\delta_{m_n-m_s+m_u-m_d,0} \times\\
&\ket{\sigma_{j,x}^z=\omega^{m_s}}
\otimes\ket{\sigma_{j,y}^z=\omega^{m_e}}
\otimes\ket{\sigma_{j,z}^z=\omega^{m_u}}.
\end{split}
\end{equation}
Notice that one can perform a basis transformation $\sigma_{j, x(y)(-z)}^z \to \sigma_{j, x(y)(-z)}^{z\dag}$ for $j$ in a bipartite sublattice to make the vertex rule, i.e. the star stabilizers in Fig.~\ref{fig:XCtensor}a, symmetric under mirror reflection $\delta_{m_w+m_e+m_u+m_d,0}\delta_{m_n+m_s+m_u+m_d,0}$. It is equivalent to doubling the unit-cell for the Hamiltonian, while the wave function is still translationally invariant. Then we get a mirror symmetric real rank-6 tensor of bond dimension $N$. 
For instance, for the simplest $\mathbb{Z}_2$ scenario there are, due to the two constraints in each vertex, $2^{6-2}=16$ allowed classical vertex configurations, which constitute the elementary vertices for the $\sigma^z$ classical configurations in the wave function, expanded as a cuboid condensate (Fig.~\ref{fig:XCtensor}fh).
\end{itemize}

Working with with the representations above, we can contract out the physical variables for all three representations of the same partition function defined by $\norm{ e^{t \sigma^x/2}\ket{\psi_0}}^2$, see Fig.~\ref{fig:XCtensor}g. The structures are the same as the wave-functions, but the tuning parameter $h$ acting upon the physical variables is elevated to control the interaction strengths between the virtual variables. 
In the $\sigma^x$ representation, it is straightforward to contract out the physical variables while sealing the ket hyper-layer and the bra hyper-layer into an extremely anisotropic Ashkin-Teller model on the cubic lattice~\cite{Johnston14PIM, Vijay17coupledLayersXcube}, which is also related to the dual height representation for the tensor gauge theory~\cite{XuWu08RVP}. There are two classical clock spins on each site, denoted as black dots in the figure, and there is an oriented clock interaction between nearest neighbors, denoted as open circle in the figure. Thus the classical model may be written as
\begin{equation}
\epsilon_{d} = -t\ \text{Re} \sum_{j}  \omega^{m_{j+x}-m_j} + \omega^{n_{j+y}-n_j} + \omega^{m_j+n_j - m_{j+z}-n_{j+z}}.
\label{eq:AshkinTellerModel}
\end{equation}
Physically speaking, this oriented interaction originates from the fact that the magnetic lines in the XC are pulled by the monopoles and therefore become one dimensional arrays. 
In the $\sigma^z$ representation, we can first go to the dual vertex representation as we have done in the previous section
\begin{equation}
\begin{split}
\langle\psi|\psi\rangle &= \bra{\psi_0} e^{\frac{t}{2}\sum_l (\sigma^x_l + \sigma_l^{x \dag})}  \ket{\psi_0}\\
&= \sum_{\{m\}} \prod_l p_{m_l}^2  \bra{\psi_0}  (\sigma_l^x)^{m_l} \ket{\psi_0} 
= \text{tTr}\prod_j \hat{T}_p(j) \,,
\end{split}
\end{equation}
which we can alternatively interpret as a vertex model or a classical TN.
The vertex constraint can be automatically satisfied by a vertex-cube duality and representing the vertex leg variable as the plaquette variable, leading to the plaquette clock model 
\begin{equation}
\epsilon_p = -\sum_{\square} \ln \left(\frac{1}{N}\sum_{k=0}^{N-1} e^{t \cos{\frac{2\pi k}{N}}} W_\square^{\prime k}\right).
\label{eq:SMplaquetteModel}
\end{equation}
The relation between the three equivalent classical models is shown in Fig.~\ref{fig:XCtensor}g. 
We note that this treatment from Eq.~\eqref{eq:AshkinTellerModel} to Eq.~\eqref{eq:SMplaquetteModel} is essentially the Kramers-Wannier duality in classical statistical physics, and indeed $-\ln p_m^2(t)$ is related to $t$ by the Kramers-Wannier relation in swapping high temperature with low temperature: the smaller coupling $t$, the larger relative weight $p_0^2(t)$ favoring the ordered phase. At the extreme $t=0$, $p_m^2(t)=\delta_{m,0}$ such that only $\{m=0\}$ configuration contributes to the partition function. 
Check that for $N=2$, $p_0^2(t)=\cosh{t}, p_1^2(t)=\sinh{t}$ which gives the dual coupling strength $t'=-\frac{1}{2}\ln (p_1^2 / p_0^2)=\frac{1}{2} \ln\coth{t}$, consistent with the well-known Kramers-Wannier relation $e^{-2t'}=\tanh{t}$ in the Ising case. 


\section{A map from $\mathbb{Z}_N$ X-cube model to $\mathbb{Z}_N$ tensor-gauge theory}
\begin{figure}[ht]
    \centering
    \includegraphics[width=\columnwidth]{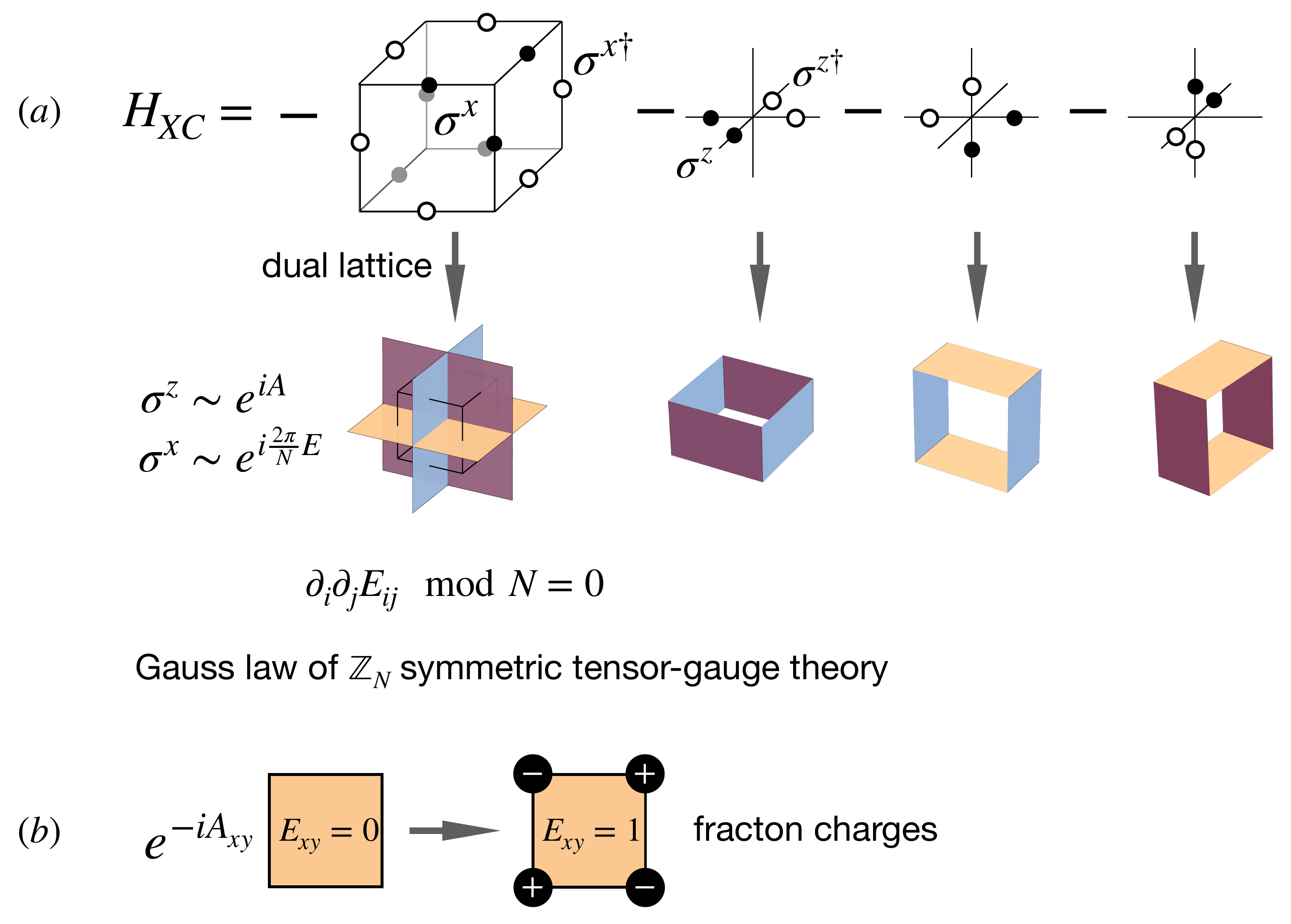}
    \caption{
    {\bf A map from X-cube model to tensor-gauge theory.}
    In the dual lattice, the field lies at the plaquette center. 
    (a) In the Hamiltonian, the cube term stabilizer is mapped to an energy penalty that enforces the Gauss law of the tensor-gauge theory surrounding a vertex i.e. setting the bare mass of the gauge charge. The star term is also mapped to some closed surfaces that sets the bare mass of the magnetic vector monopole. 
    (b) Increasing electric field in the plaquette center creates the fracton gauge charges at the corners. 
    }
    \label{fig:map2gauge}
\end{figure}

As shown in Fig.~\ref{fig:map2gauge}a, the $\mathbb{Z}_N$ Pauli matrices on the original link center can be mapped to the electric field $E$ and gauge field $A$ on the dual plaquette center. In this way, the X cube stabilizer term enforces the Gauss law 
\begin{equation}
\partial_i\partial_j E_{ij} \mod N= 0
\end{equation}
of $\mathbb{Z}_N$ tensor-gauge theory in the ground state. In this case, the diagonal element of the symmetric electric field tensor (a 3-by-3 matrix) is restricted to zero, and the off-diagonal element e.g. $E_{xy}$ resides on the plaquette center in $xy$ plane. This is described by the hollow tensor gauge theory~\cite{XuWu08RVP, Chen18tensorhiggs,Seiberg21ZNfracton}. Such gauge theory naturally emerges in gauging a matter with subsystem symmetry, which corresponds to subsystem charge conservation law, a stronger version of the dipole conservation law. 
In Fig.~\ref{fig:map2gauge}b, as the gauge field $A_{xy}$ is a canonical variable with $E_{xy}$, applying the generator $e^{-iA_{xy}}$ onto a plaquette raises $E_{xy}$ by one unit, and creates four gauge charges surrounding the plaquette, consistent with the Gauss law. Notice that the total charges in any plane is conserved in this process. 
On top of the vacuum, such Gauss law requires that all the gauge charges lie at the corner of certain membrane with nonzero electric field. Moving the gauge charges is equivalent to moving the electric field membrane, which explains the origin of fracton mobility: a corner of a membrane cannot move alone by itself. 

Such gauge description can be generalized to the compact $U(1)$ case, when $A$ becomes continuous and $E$ is unlimited integer, and the Gauss law no longer needs to be $\mod N$. The $\mod N$ part in the Gauss law actually corresponds the physical Higgs mechanism: one can take the $U(1)$ model as a parent state and apply Higgs mechanism to condense $N$ charges into any finite $\mathbb{Z}_N$ model. 
Alternatively, one can also start from the discrete $\mathbb{Z}_N$ model and increase $N$ to approach the unHiggsed compact $U(1)$ limit. 

\section{Boundary fixed point iPEPS method}
In this section we elaborate on the numerical contraction of the 3D TN using the boundary fixed point iPEPS method, and discuss its relation to the reduced density matrix and second order R\'enyi entropy in a half-system partition. 

\begin{figure*}[ht]
    \centering
    \includegraphics[width=1.8\columnwidth]{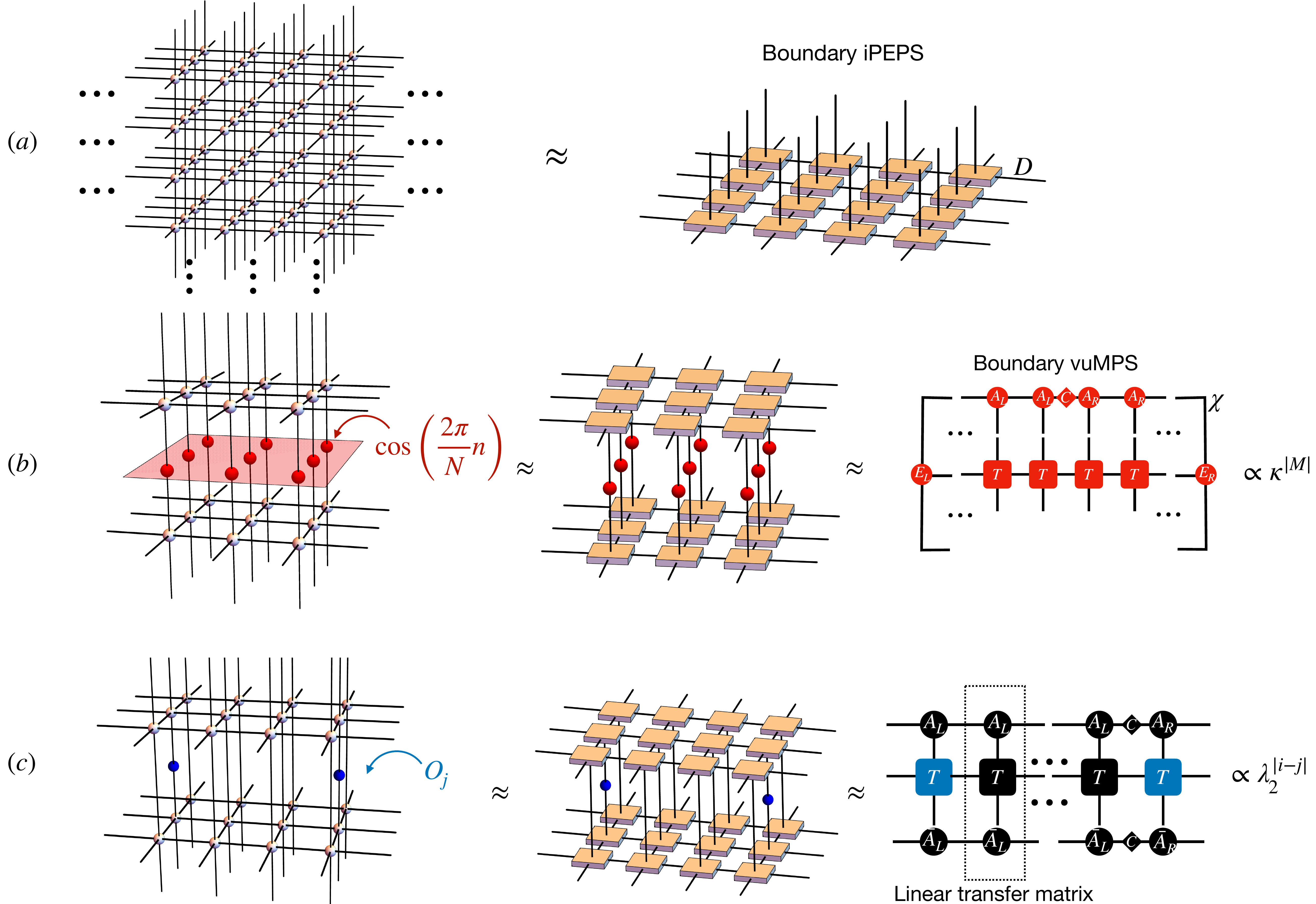}
    \caption{
    {\bf Tensor network calculation for physical observables.}
    (a) Semi-infinite classical TN with dangling bonds in a 2D interface is captured by its fixed point, which can be approximated by a variational uniform iPEPS of bond dimension $D$.
    (b) The membrane correlation is defined as the expectation value of a product of local diagonal operators $(Z+Z^\dag)/2=\oplus_{n=0}^{N-1} \cos(\frac{2\pi}{N} n)$ over the red membrane being inserted into the bonds of the TN of $\langle \psi|\psi \rangle$. 
    The resulting 3D TN is approximately compressed by iPEPS method into a two dimensional TN, which is further compressed into a one-dimensional MPS of bond dimension $\chi$. The iPEPS and the vuMPS are renormalized. 
    For infinitely large membrane, the asymptotic membrane correlation length scale can be calculated by $\sqrt{-1/\ln{\kappa}}$ where $\kappa$ is the unit-cell fraction of the number by fully contracting the 2D TN.  
    (c) 
    The two point correlation or one point observable can be likewise computed by inserting the corresponding operator into the virtual bond of the TN. For example, inserting a single $\cos{(2\pi n/N)}$ operator measures the $\mathbb{Z}_N$ spin magnetization Re$\langle \psi | \mu^z | \psi \rangle$; inserting a single $X$ operator measures the norm of a charge excitation $\langle e|e\rangle$.
    The correlation length $\xi=-1/\ln{\lambda_2}$ can be computed from the sub-leading eigenvalue of the effective linear transfer matrix outlined by the dashed line. 
    }
    \label{fig:TNCal}
\end{figure*}

To efficiently contract the 3D tensor network numerically, we employ the variational iPEPS method for classical statistical models~\cite{Verstraete18classic3D}. 
Each slice of the 3D TN is a tensor product operator and can be viewed as a transfer matrix. Its dominant eigenvector represents the semi-infinite large subsystem, often called a \textit{fixed point}, which can be approximately parametrized as an iPEPS with moderate finite bond dimension $D$. The effective free energy of the classical TN, i.e.\ $-\ln{\langle \psi | \psi \rangle}$ (see Table I of the main text), as well as its gradient with respect to the iPEPS tensor element, can be obtained by contracting out the transfer tensor product operator sandwiched by the iPEPS ansatz. The effective free energy is iteratively minimized by the quasi-Newton (LBFGS) algorithm carried out in our Julia implementation with the NLopt package \cite{Wright06optimBook, NLoptPackage} until the norm of the gradient tensor (absolute maximum of the tensor elements) drops below the tolerance threshold (usually taken as $\lesssim 10^{-6}$ in our case). Each iteration requires contracting out two 2D TNs, which can be carried out using uniform matrix product state (MPS)  to approximate fixed point of 2D transfer operator~\cite{Verstraete16gradientmethod}, as shown in Fig.~\ref{fig:TNCal}. In the end, one arrives at a self-consistent iPEPS as the boundary fixed point. 

The numerical approximation is controlled by the virtual bond dimension of the iPEPS denoted as $D$ and the virtual bond dimension of the vuMPS denoted as $\chi$. 
Generally, the iPEPS calculation is by default set in an infinite system size, but the correlation length it can reach may be bounded by the PEPS bond dimension, which leads to the finite bond dimension effect. There can be different ways of finite bond dimension scaling in the literature. (But we also caution the reader that a PEPS with finite bond dimension does not necessarily have finite correlation length, because  counterexamples can easily be  constructed by mapping any two dimensional discrete classical model into the coherent quantum wave function~\cite{Henley04RK, Fradkin04RK}).
For each iteration during the numerical optimization we need to contract specifically a 2D TN with total bond dimension $ND^2$ for $\langle \phi | \hat{T}|\phi\rangle$ and a 2D TN with bond dimension $D^2$ for $\langle \phi |\phi\rangle$, where $|\phi\rangle$ denotes the  fictitious 2D quantum wave function generated by the iPEPS. 
Notice that the 2D transfer tensor product operator may be interpreted as a imaginary time evolution operator of a 2D quantum system, and the optimization for the boundary fixed point is analogous to finding the 2D quantum ground state as an infinite PEPS~\cite{Verstraete16gradientmethod, Corboz16ipeps, Xiang19ADtensor, Corboz18, Lauchli18, Chen20su3peps}. The difference is in that in our case we need to optimize a rank-6 tensor product operator $T$ instead of a local rank-2 Hamiltonian matrix, and so the computation complexity is higher ($ND^2$ vs. $D^2$).
Further virtual symmetries inside the iPEPS, like the ones considered in Ref.~\cite{Schuch21entanglementorder},are not explicitly considered here, because we want to fully utilize the limited finite $D$ space we can get especially for large $N \gtrsim 10$. Also it does not harm because the entanglement symmetry of our 3D quantum TN state acts on the physical indices of the boundary iPEPS instead of the PEPS virtual dimension. Nevertheless, we always impose a mirror reflection symmetry and fix ourselves a priori to the trivial representation i.e.\ no sign change upon mirror reflection, required by the vuMPS method for good performance. 
 Unlike solving a 2D quantum Hamiltonian, the iPEPS $|\phi\rangle$ we optimize is the boundary fixed point that contains the information for the reduced density matrix when half of the system in the transfer direction is traced out~\cite{Verstraete13toriccode}. 
Let us denote the optimized boundary fixed point iPEPS as 
\begin{equation}
    |\phi\rangle = \text{tTr}\big(\prod_j A^{n}(j)\big) \,,
\end{equation}
where $A^n(j)$ is a rank-5 PEPS tensor at site $j$ with a dangling bond $n=0,\ldots,N-1$ corresponding to the virtual bond of the 3D TN. One should not confuse this rank-5 PEPS tensor with the boundary MPS rank-3 tensor drawn in Fig.~\ref{fig:TNCal}.

\begin{figure}[ht]
    \centering
    \includegraphics[width=\columnwidth]{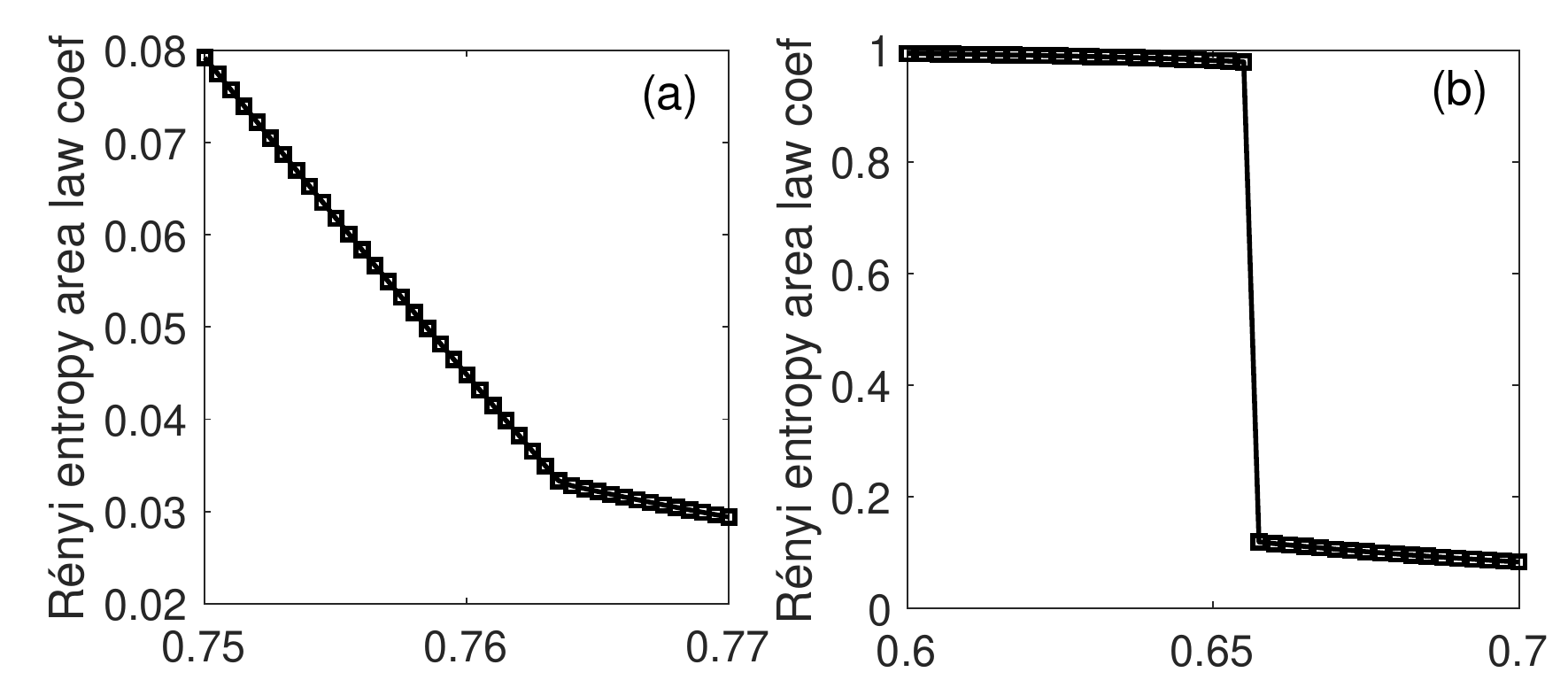}
    \caption{{\bf Area law coefficient} $a$ for the 2nd Rényi entropy $S_R = a L^2\ln{N} + \ldots$ for the QPT (a) between $\mathbb{Z}_2$ 3D toric code state and trivial state; (b) between $\mathbb{Z}_2$ 3D X cube state and trivial state. $L$ is the linear system size. It is based on the $D=2$ data. }
    \label{fig:Z2gaugeEnt}
\end{figure}

For $\hat{T}_g$, the classical gauge model, the TN is obtained from the $m$-loop condensation phase transition, recall that the virtual bond is compressed from the double layer TN $\langle\psi|\psi\rangle$, where the virtual indices from the bra layer and ket layer are identical due to the diagonal deformation. The reduced density matrix of tracing out half of the infinitely large cubic lattice in open boundary becomes a diagonal matrix generated by the uniform tensor product operator:
\begin{equation}
    \hat{\rho}_{\{n\},\{n\}}= \text{tTr}\left(\prod_j A^n\otimes A^n(j)\right),
\end{equation}
where $n_j$ is left as the dangling indices representing the effective boundary degree of freedom. The normalization of reduced density matrix is equivalent to the normalization of the iPEPS $\text{Tr}\hat{\rho} = \langle \phi|\phi\rangle$. 
We can then derive the second order R\'enyi entropy by
\begin{equation}
\begin{split}
    &e^{-S_{R}} = \text{Tr}\hat{\rho}^2\\
    &=\text{tTr}\left(\prod_j \sum_{n}A^n\otimes A^n\otimes A^n\otimes A^n(j)\right) 
    \equiv e^{- a L^2\ln{N}},
\end{split}
\end{equation}
in which the contraction of four-layer 2D TN can again be  computed by vuMPS method for the area law coefficient $a$ of the entanglement entropy. Notice that the trace of the second order reduced density matrix here resembles the inverse participation ratio of $|\phi\rangle$ in the many-body space. The result is shown in Fig.~\ref{fig:Z2gaugeEnt}a, where a kink appears near the phase transition point $h_c\approx 0.7614$. 

For $\hat{T}_p$, the classical plaquette model responsible for fracton phase transition, 
this is already a compressed TN for $\langle \psi|\psi \rangle$, for which we need to retrieve the discrimination between the ket layer and the bra layer. Recall that the deformation is off-diagonal by $\sigma^x$, and so the compressed virtual bond dimension is taken as the offset between the ket layer and bra layer.
The reduced density matrix follows as
\begin{equation}
    \hat{\rho}_{\{m\},\{m'\}}= \text{tTr}\left(\prod_j \frac{1}{N}\sum_{m''}A^{m-m''}\otimes A^{m'-m''}(j)\right) \,,
\end{equation}
which is a double layer tensor product operator. The normalization of the reduced density operator is then equivalent to the normalization of the iPEPS $\text{Tr}\hat{\rho} = \langle \phi|\phi\rangle$. Again the contraction of the four-layer 2D TN can be computed by the vuMPS method for the area law coefficient $a$ of the entanglement entropy, see Fig.~\ref{fig:Z2gaugeEnt}b. 
At $t=0$, the fixed point becomes exactly a polarized state such that $A^{m\neq0}=0$, such that $\hat{\rho}$ becomes a uniform diagonal matrix. In an infinite lattice under open boundary condition, there is no global constraint. Therefore $\hat{\rho}$ is an infinite temperature density matrix supported at the boundary, yielding the maximal area-law entropy: $S_R=L^2\ln N+\ldots$.

\section{Supplemental data for $\mathbb{Z}_N$ toric code QPT}

The 3D toric code QPT also shows a kink for the magnetization~\cite{Trebst07toriccode} and a peak for the correlation length, due to finite bond dimension effect. In principle the correlation peak shall diverge if larger bond dimension is used to further optimize the boundary PEPS~\cite{Lauchli18, Corboz18, Verstraete18classic3D,Schuch21entanglementorder}. 
Notice that for $\mathbb{Z}_2$ case, $D=4$ does not gain significantly more free energy in our practice within accessible computation runtime, and so does not lead to significantly different observables than $D=3$ case, which is also the practically economically optimal bond dimension for 3D Ising model in Ref.~\cite{Lauchli18, Verstraete18classic3D, Schuch21entanglementorder}. 
We present the data for $(N=2,D=3,\chi=72)$ and $(N=5,D=2,\chi=80)$ in the main text. 

\begin{figure}[ht]
    \centering
    \includegraphics[width=\columnwidth]{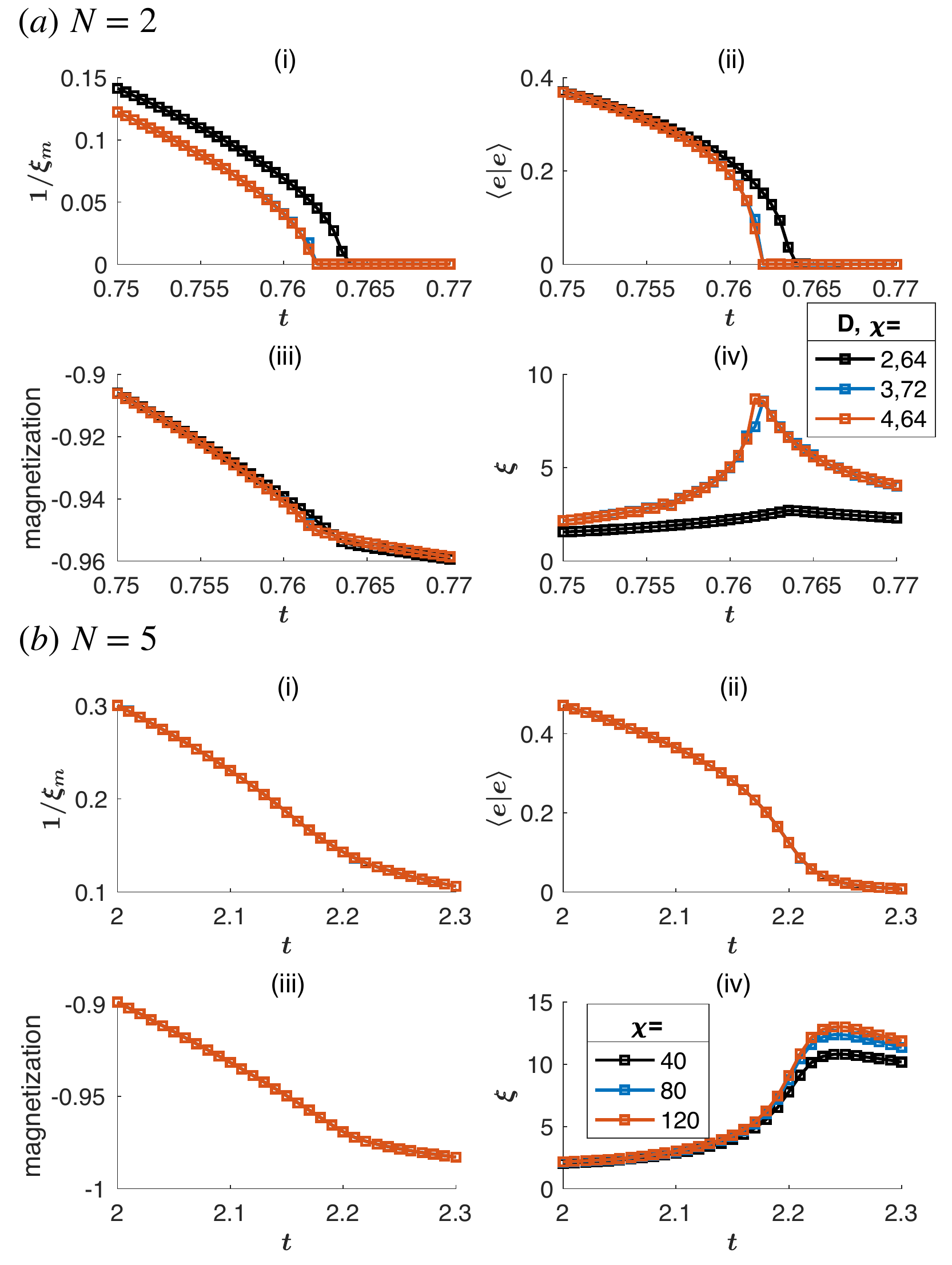}
    \caption{{\bf Supplemental observables for 3D $\mathbb{Z}_N$ toric code QPT} for (a) $N=2$ and (b) $N=5$ for varying bond dimensions $D,\chi$.}
    \label{fig:corrlgth}
\end{figure}


\section{Supplemental data for $\mathbb{Z}_N$ X cube QPT}

Here we show additional data for the $\mathbb{Z}_N$ X cube confinement transition based on the plaquette model calculated using different iPEPS dimensions $D$ and vuMPS dimensions $\chi$. The main observation is that the finite bond dimension effect is rather weak for finite $N$, due to the first-order nature of the transition. 

\begin{figure*}[t]
    \centering
    \includegraphics[width=\textwidth]{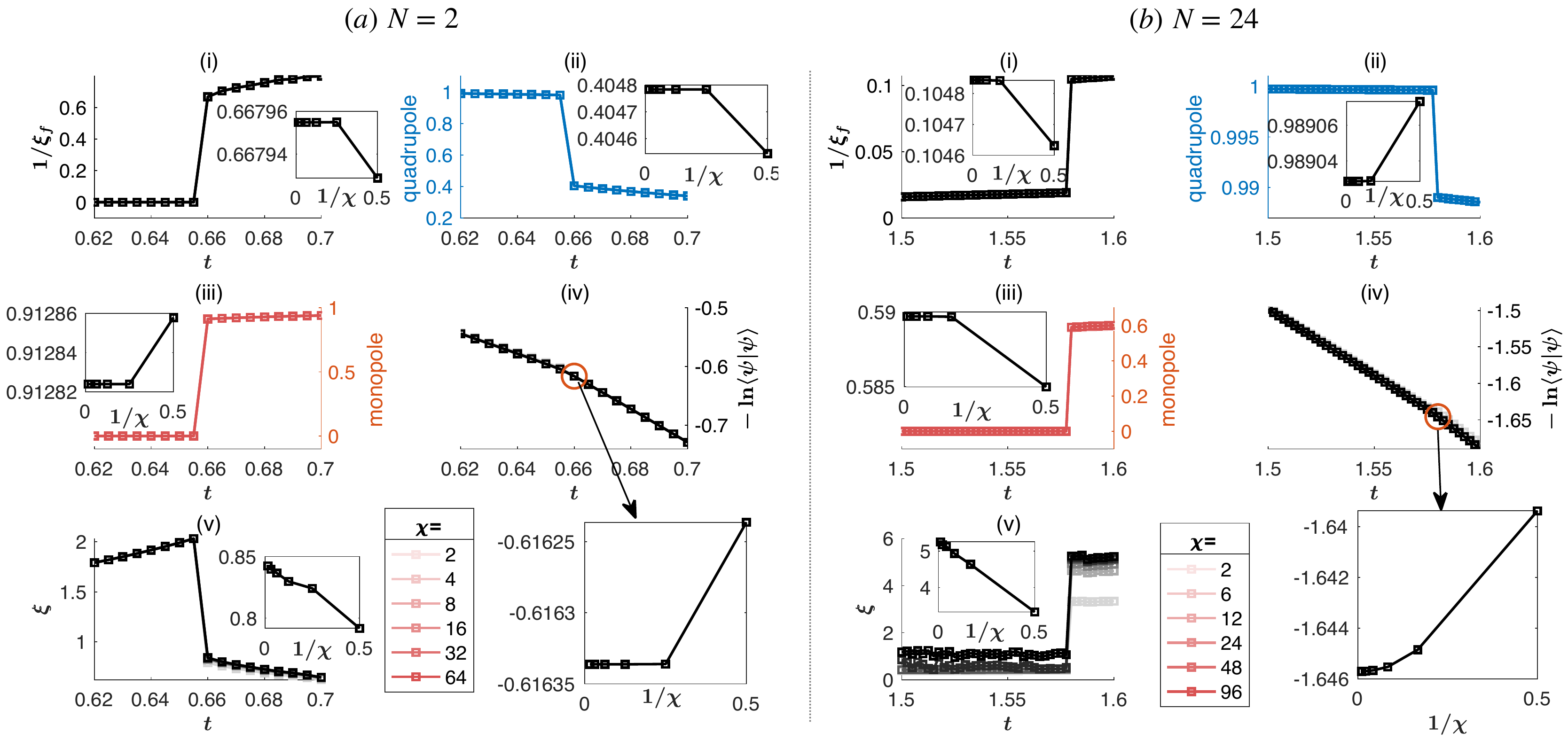}
    \caption{{\bf Finite MPS dimension effect} shown for the (a) $Z_{2}$ (left panel) and (b) $Z_{24}$ (right panel) plaquette models with fixed PEPS bond dimension $D=2$ but varying MPS bond dimension $\chi$.
    The insets show the corresponding observables at $t_c^+$ i.e. the critical point at the disordered confined phase side, versus the inverse MPS bon dimension $1/\chi$. 
     For the $\mathbb{Z}_2$ scenario, it indicates that $\chi=4$ is already approximately sufficient in yielding the converged observables except the correlation length, which can further grow for larger $\chi$. 
     For the $\mathbb{Z}_{24}$ scenario, larger $\chi$ up to $\chi=96$ gains more classical free energy, but most of physical observables do not change significantly with growing $\chi$, except the correlation length. }
    \label{fig:chiScaling}
\end{figure*}

\begin{figure*}[t]
    \centering
    \includegraphics[width=\textwidth]{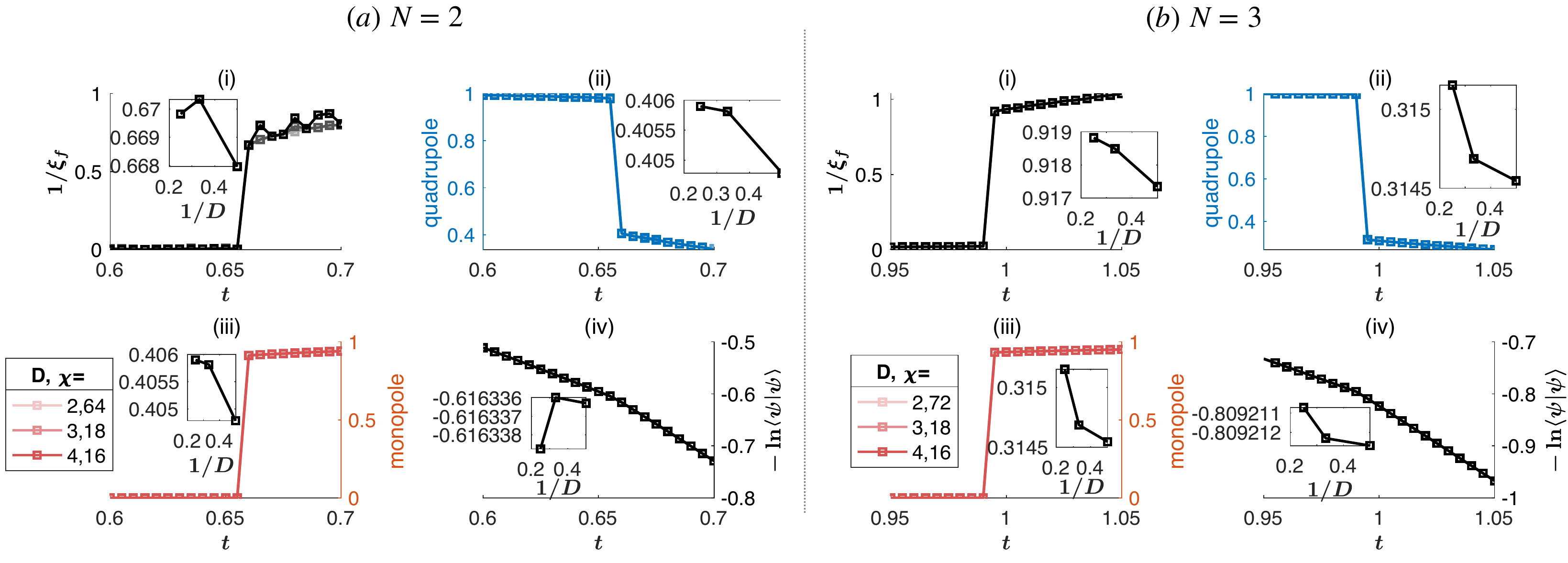}
    \caption{{\bf Finite iPEPS dimension effect} for the (a) $\mathbb{Z}_2$ (left panel) and (b) $\mathbb{Z}_3$ (right panel) plaquette model. Insets show the $D$ dependence of observables at $t_c^+$. }
    \label{fig:ZNDscaling}
\end{figure*}

As summarized in Fig.~\ref{fig:chiScaling}, we explore MPS bond dimension effects by employing the iPEPS algorithm with fixed $D=2$ and then optimizing for vaying $\chi$. The calculated observables for varying $\chi$ do not significantly change, except for the classical correlation length. Then for small $N=2,3$, we can also upgrade the optimization for larger iPEPS bond dimensions $D\to 3,4$, as shown in Fig.~\ref{fig:ZNDscaling}. In practice, we have run the optimization with different starting points, using both  random symmetric initial points as well as converged numerical samples of smaller bond dimension $D=2$ decorated with symmetric noise of order $10^{-3}, 10^{-4}$. However, despite several such runs for each parameter point, we find it rather difficult to further lower the free energy by more than $0.0001$. This situation is quite unlike the gauge model with its continuous phase transition, where, for the Ising case, $D=3$ yields significant improvements over $D=2$ as shown in Fig.~\ref{fig:corrlgth}. We attribute this numerical phenomenon to the strong first-order nature of the  transition for small $N$ in the plaquette model, which therefore has short correlation lengths and can be sufficiently described by iPEPS with only small bond dimensions. 

In the following we comment two technical issues related with local minimum in the optimization process. 

Firstly, near the critical point $t\simeq t_c$, due to the first order transition nature, there is a significant adjacent metastable local minimum, which could trap the iPEPS optimization leading to the hysteresis-like pattern near the kink in the free energy curve. Hence we take at least two limiting samples for optimization, one starting from a fully random real mirror symmetric iPEPS i.e. disordered state, and one starting from a strongly polarized random real mirror symmetric iPEPS, where we multiply a factor of $\simeq 10^{-2}$ to the physical indices $n(m)>0$. Then we compare their final free energy densities. Alternatively, if time allows one may also sweep the phase diagram sequentially from left to right and from right to left, using the previous converged iPEPS as the starting point for next parameter, which shall display a hysteresis loop. 

Secondly, in the ordered phase for $t\lesssim t_c$, the boundary iPEPS is a state very close to a polarized state, which is a numerical singularity point that can also trap our optimization process. 
For $\mathbb{Z}_2$, with the diagram for the 16 vertices explicitly drawn in Fig.~\ref{fig:XCtensor}h, with each leg weighted by $p_m(t)$, we can gain some analytical insight into the boundary iPEPS as the dominant eigenstate for the linear transfer map generated by a layer of vertices. Deep in the classical ordered phase $t\ll t_c$, the vertex leg weight function $p_1\ll p_0$, so that the vacuum configuration $\{m=0\}$ dominates, which allows a low-temperature expansion into this linear transfer operator:
\begin{equation}
\includegraphics[width=\columnwidth]{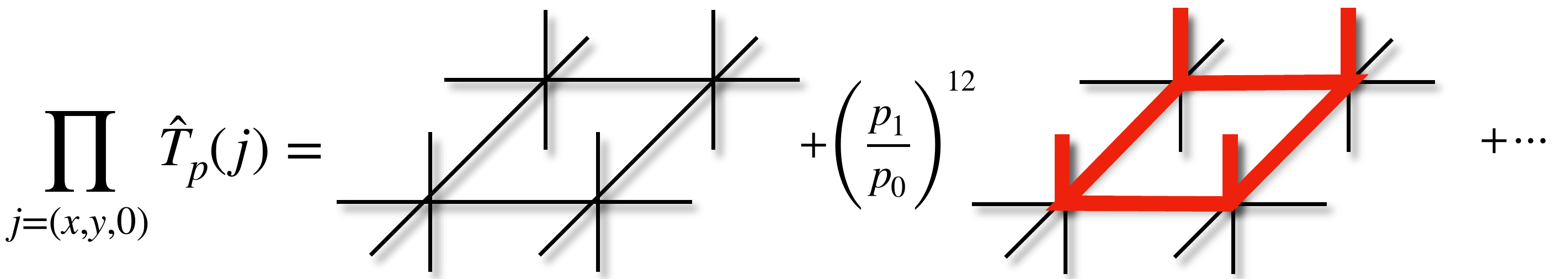},
\label{eq:plaqTrsfOpExp}
\end{equation}
where the subleading term is the fluctuation of a minimal half-cuboid shifting four nearest neighbor variables $0000\to 1111$. The fluctuation with larger cuboid scales exponentially with its total edge lengths reminiscent of the perimeter-law of Wilson loops in the classical gauge theory. But compared with loop fluctuations in the vector gauge theory, the cuboid fluctuations generally invoke more legs and take the form of higher order perturbations. 
Therefore, away from the ordered limit $t=0$ where $\{m=0\}$ and $W_\square'=1$, we would expect relatively weak fluctuations to $W_\square'\lesssim1$ which is the quadrupole amplitude we show in Fig.~3bd of the main text. Given the fact that the correlation length does not need to grow too large towards the first order phase transition, the feature of weak fluctuation may persist until the the transition. Indeed this is what we observe in our iPEPS numerical computations. 
Note that the boundary iPEPS for the ordered phase can thus be well approximated by a simple fully polarized state, i.e.\ a mean-field ansatz. 

One should note, however, that while the fully polarized state might provide the cheapest numerical approximation to the boundary, it can also become a dangerous trap for the iPEPS optimization based on a single-site-gradient-tensor. The latter is obtained by sandwiching the transfer operator with the variational PEPS leaving only one vacancy. In this case, it is obtained by fixing all the top and bottom indices of the linear transfer operator into 0 while leaving only one-site vacancy:
\begin{equation}
\includegraphics[width=\columnwidth]{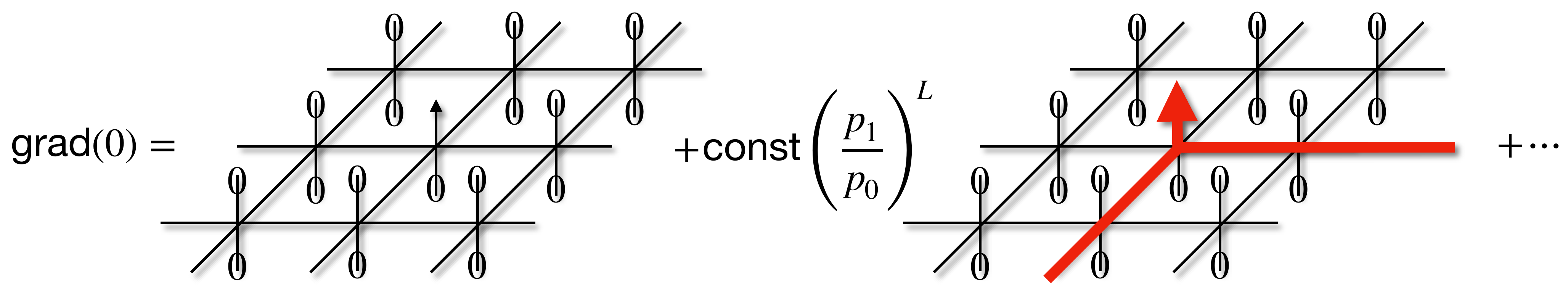}.
\end{equation}
From the vertex rule one can deduce that the single-site gradient tensor, in an infinite lattice, would have exponentially small components away from the polarized limit, prohibiting further optimization search. 
In other words, the fully polarized state is a singular point in the numerical computation, which is not a true physical local minimum. This numerical singularity in the single-site gradient originates from the strict elementary vertex rule in the linear transfer map, which also occurs in the gauge model. Generally it would not become a practical problem when the true global optimized iPEPS is far away from this singular point. 
However, in the ordered phase of the plaquette model, it is actually quite close. In practice, we therefore mainly use random symmetric iPEPS as the starting point for the first round of optimization. Then to avoid being trapped in the physical local minimum due to the first order transition, we run another optimization using a random symmetric iPEPS with $m>0$ component suppressed by a factor of $10^{-2}$ as our starting point for optimization, which would favor the ordered phase. Occasionally we also run several more optimizations with different random starting iPEPS. In the end, we keep only the one convergent numerical sample with lowest free energy.
As it turns out, near $t\lesssim t_c$, some of our numerical optimized iPEPS samples are indeed attracted into the polarized trap vicinity, evaluating the diagonal onsite observable extremely close to 1.0. Under this situation we would add some noise of order $10^{-1}$ to iPEPS and run the optimization again, until we get to a new minimum, which sometimes gains $\sim 0.1\%$ more free energy density than the polarized state.
Moreover, we find that the singular case of the $D=1$ product state can only approximately describe the ordered phase $t<t_c$, but not the disordered confinement phase $t>t_c$. For $t>t_c$ we have numerically confirmed that the classical free energy of the disordered local minimum for the $D=1$ ansatz is higher than the ordered local minimum. 
Last but not least, we also comment that the fact that the boundary iPEPS fixed point becomes an exact polarized state at $t=0$ does not imply trivial physics in the 3D quantum wave function, an XC ground state at its fixed point. In fact, the boundary iPEPS captures the reduced density operator. The polarized boundary PEPS in this basis yields diverging fracton confinement lengthscale, although the monopole correlation length is zero, and it leads to a reduced density matrix with maximal area-law entropy. 


\section{Fracton dipole condensation in pure X cube}
Besides the fracton confinement transition tuned by $t\sigma^x$, here we also briefly comment on the fracton dipole condensation transition tuned by $\lambda\sigma^z$. 
$\sigma^z$ acting on a link creates the tightest fracton quadrupole on the adjacent cubes, and fluctuates the tightest fracton dipole, which is a planar composite particle mobile in the plane perpendicular to the dipole bond. Its condensation criticality is thus argued to be described by a coupled 2D CFT~\cite{Hermele21fracton}. 
In our wave-function deformation approach, where we use a TN representation in the $\sigma^z$ basis as shown in Fig.~\ref{fig:XCtensor}, it is straightforward to see that it is directly described by a plaquette model controlled by $\lambda$ acting as inverse temperature
\begin{equation}
\bra{\psi_0} e^{\frac{\lambda}{2}\sum_l  \sigma_l^z+h.c.} \ket{\psi_0}
= 
\sum_{\{m\}} e^{\lambda \sum_\square \text{Re} W_\square'}.
\end{equation}
For the $\mathbb{Z}_2$ scenario we can immediately deduce a first-order transition at $\lambda_c = \frac{1}{2}\ln\coth{t_c}$, that separates the XC, from the fracton dipole condensed trivial paramagnet phase. 
For the generic $N$ scenario, it is not exactly equivalent to the model we computed, because there is no higher power of plaquette interaction in this case. 
Cast in the vertex representation, we have the same vertex rules, but the leg weight function $p_n$ for $n=0,1,\ldots,N-1$ is quantitatively modified. 
The numerical computation for this model remains to be explored in future work.


\section{Comparison with previous studies into the fracton QPT}
The XC topological order is essentially a lattice $\mathbb{Z}_N$ hollow(off-diagonal) tensor gauge theory coupled with fractonic matter charge. 
Ref.~\cite{Hermele21fracton} recently discusses the low energy effective field theories for its QPT: while a single fracton charge is too restricted to fluctuate, the authors mainly discussed the matter fluctuation of fracton electric charge dipole, and the gauge fluctuation of a lineon magnetic monopole (or monopole dipole), in a separate manner. By condensing different subsets of those electric or magnetic particles with restricted mobility along a line or within a plane, the authors phenomenologically map the problem into the problem of stacking 2D CFT or stacking 1D CFT wires, subject to generic coupling. They discuss the relevance of the inter-CFT coupling to draw conclusion for the stability of the critical points or gapless intermediate phases, qualitatively consistent with earlier Monte Carlo calculations in the Appendix of Ref.~\cite{Kim17}. 
A more recent Ref.~\cite{Williamson21SubdimensionalCriticality} also performs systematic study into the subdimensional criticality by imposing the organizing principle of subsystem symmetry breaking, which can describe various types of quantum phase transitions involving fracton physics. 
The confinement transition of $\mathbb{Z}_2$ XC was numerically confirmed to be a first-order transition in the stochastic series expansion (SSE) calculations as shown in supplemental material in Ref.~\cite{Sondhi18fractondiagnostics}.
A similar conclusion was also drawn by a perturbative analytical study using the perturbative-continuous-unitary-transformation method in Fig.~12 and Fig.~27 of Ref.~\cite{Schmidt20fractonpcut}, where the authors performed series expansion up to about eighth order from the deconfined limit and the confined limit, for the ground state energy as well as the monopole mass gap in the phase diagram. 
The analytic perturbative approach was also employed in the Ref.~\cite{Liu21fracton} to pin the location of transition point. 
In a more recent paper the authors also use the same series expansion method to study a phase diagram of directly interpolating the Hamiltonian between the 3D toric code and X cube model~\cite{Walther22Xcube}, by tuning the coupling constant of the XC inplane star and cube stabilizer terms from a TC phase. Although their phase diagram also consists of both 3D TC and XC phases, bearing certain similarity to our phase diagram, the nature of their phase transitions are qualitatively sharply distinct from ours: by directly tuning the inplane star stabilizer terms $\hat{A}_+$, the authors therein are essentially tuning the rest mass of a monopole in XC phase; likewise, tuning the cube stabilizer terms $\prod_{l\in\ \mbox{\mancube}}\sigma_{l}^x$ is equivalent to tuning the rest mass of a fracton defect. This view also holds near the TC phase if one interpret the electric string turning point as a monopole, and the resonant state of an electric charge octupole around an elementary cube satisfying $\prod_{l\in\ \mbox{\mancube}}\sigma_{l}^x=-1$ as a fracton. Viewed in the basis of the topological defects, such tuning parameter couples to the onsite diagonal mass for a given single topological defect, which explains the first order transition found by the authors~\cite{Walther22Xcube}. In contrast, the gauge fluctuations are interactions between topological defects at different locations, such as the pair annihilation of the monopoles from adjacent vertices, or the loop fluctuation of the magnetic flux penetrating different faces. 

To summarize, in this paper we mainly discuss the confinement transition due to pure gauge fluctuations: the vector-gauge magnetic flux loop fluctuation together with the tensor-gauge magnetic monopole fluctuation. Compared with the previous Hamiltonian study for the confinement transition, our paper highlights the wave-function approach with 3D spatial conformal quantum critical points. Moreover, we generalize the study to larger $N$ and embed the fracton order in the coupled layer construction of 2D topological order. The relation between the Hamiltonian QPT and the wave-function QPT has been studied for the 2D toric code scenario~\cite{Isakov2011}, but it remains to be explored for the 3D fracton scenario when going to large $N$, for which the continuous field theory at the gapped fixed point is already rather nonstandard let alone the critical point. Our exact 3D iPEPS wave function serves as a good starting point for further optimization to minimize the energy of a 3D quantum Hamiltonian~\cite{Corboz22ipeps3d}, which is a promising route to study this more exotic connection between the wave function criticality and the Hamiltonian criticality. 

\end{document}